\documentclass[12pt]{article}

\usepackage{booktabs}  
\usepackage{xcolor}

\usepackage[colorlinks,citecolor=blue,backref=page]{hyperref}
\usepackage{cite}
\usepackage{amsmath,amsfonts,amssymb}
\usepackage{float}
\usepackage{graphicx}
\usepackage{latexsym}
\usepackage{enumitem}
\usepackage{bm}

\usepackage{wrapfig}

\usepackage{amsthm}
\usepackage{xspace}

\usepackage{cleveref}
\usepackage{array}
\usepackage[normalem]{ulem}
\usepackage{comment}
\usepackage{bbm}

\usepackage{scalerel}
\usepackage{mathtools}
\usepackage{cases}
\usepackage{relsize}

\input{ti_evenfewermacros.sty}

\usepackage[margin=1.5in]{geometry}

\newcommand{\uvector}{\mathbf{e}}

\newcommand{\kOtheta}{k_{0\theta}}

\newcommand{\zd}{\z_\text{d}}
\newcommand{\td}{t_\text{d}}

\newcommand{\xid}{\xi_\text{d}}
\newcommand{\zetad}{\zeta_\text{d}}
\newcommand{\etad}{\eta_\text{d}}
\newcommand{\psid}{\psi_\text{d}}
\newcommand{\zdO}{z_{\text{d}1}}
\newcommand{\zdT}{z_{\text{d}2}}
\newcommand{\nuz}{\nu_\text{b}}
\newcommand{\nut}{\nu_t}
\newcommand{\nus}{\nu_s}

\newcommand{\muz}{\mu_\text{b}}
\newcommand{\mut}{\mu_t}
\newcommand{\mus}{\mu_s}
\newcommand{\zetamax}{\zeta_{\max}}
\newcommand{\zetamin}{\zeta_{\min}}

\newcommand{\sigmasqz}{\sigma^2_\text{b}}
\newcommand{\sigmasqt}{\sigma^2_t}

\newcommand{\sigmasqs}{\sigma^2_s}
\newcommand{\sigmasqalpha}{\sigma^2_\alpha}
\newcommand{\sigmasqn}{\sigma^2_\text{n}}

\newcommand{\phiT}{\varphi_T}

\newcommand{\avg}[1]{\left\langle#1\right\rangle}

\newcommand{\Wlin}{W_\text{lin}}

\newcommand{\bPhi}{b_\Phi}
\newcommand{\una}{\text{U}}

\newcommand{\Phiarg}[2]{\Phi\big(\kOtheta\phiT #1, \kOtheta\phiT^2 #2\big)}

\newcommand{\roverline}[1]{\mathpalette\doroverline{#1}}
\newcommand{\doroverline}[2]{\overline{#1#2}}

\newcommand{\tagprime}[1]{\tag{\ref*{#1}$'$}}

\DeclareMathOperator{\sign}{sign}
\DeclareMathOperator{\cdf}{cdf}

\DeclareMathAlphabet{\brm}{OT1}{cmr}{bx}{rm}

\title{Statistical characterization of scattering delay in synthetic aperture radar imaging}

\author{Mikhail Gilman and Semyon Tsynkov}

\begin{document}

\maketitle

\begin{abstract}

Distinguishing between the instantaneous and delayed scatterers in synthetic aperture radar (SAR) images is important for target identification and characterization. To perform this task, one can use the autocorrelation analysis of coordinate-delay images.
However, due to the range-delay ambiguity the difference in the correlation properties between the instantaneous and delayed targets may be  small. Moreover, the reliability of discrimination is affected by speckle, which is ubiquitous in SAR images, and requires statistical treatment. 

Previously, we have developed a maximum likelihood based approach for discriminating between the instantaneous and delayed targets in SAR images. To test it, we employed simple statistical models. 
They allowed us to simulate ensembles of images that depend on various parameters, including  aperture width and  target contrast.

In the current paper, we  enhance our previously developed methodology by establishing confidence levels for the discrimination between the instantaneous and delayed scatterers. Our procedure takes into account the difference in thresholds for different target contrasts without making any assumptions about the  statistics of those contrasts. 

\end{abstract}

\newcommand{\dataR}{\mathbf{R}}
\newcommand{\gts}{g_{ts}}

%
%
%
%
%
%
%

\section{Introduction}
\label{sec:stRadio_Introduction}

Signal processing algorithms for synthetic aperture radar (SAR) imaging require a model for signal propagation  and a model for scattering about the target. For example,  standard SAR assumes dispersionless propagation with the speed of light and a point scatterer with constant  instantaneous reflectivity. These models are  deterministic. 
In addition, stochastic treatment may be justified for certain imaging scenarios.
For example, physical characteristics of a turbulent medium (e.g., density, velocity, etc.) are typically considered   random fields~\cite{monin1,rytov4}. Accordingly,  if  radar signals propagate through such a medium, the resulting SAR image is described in statistical terms \cite{meyer-2016,solna-13,a83e}. 

Stochastic approach can also be used to describe the scattering of SAR signals about the target. 
A detailed stochastic treatment of  instantaneous extended  targets can be found in~\cite{oliver-98}. 
The goal of the current study is to address the  scattering delay and its detection in SAR. 
For scatterers with delayed response, a stochastic model for SAR imaging  has been built in our  work~\cite{a85e}. It assumes the deterministic propagation with constant speed as in standard SAR, while scattering about both  instantaneous and delayed targets is described in stochastic framework. Hereafter, we extend the results of~\cite{a85e} by introducing  confidence levels for the detection of targets with delayed response. 

A delayed component in scattering may carry valuable information about the properties of the scatterer, such as its internal structure and characteristic size. The main difficulty in detecting the scattering delay   is to separate it from the propagation delay, which is at the core of  SAR reconstruction. 
The authors of \cite{ferrara-2017} propose to interpret the scattering delay as a third dimension added to  target reflectivity and SAR image (on top of two spatial coordinates). 
  Then, a  point scatterer in space $\sim \delta(\z - \z_0)$ used in  standard SAR is replaced with a  point scatterer in space-time $\sim \delta(\z-\z_0)\delta(t-t_0)$. Subsequent analysis in~\cite{ferrara-2017} focuses on the resulting coordinate-delay point spread function (PSF). The aforementioned difficulty in separating  scattering delay from the propagation delay manifests itself via slow decay of PSF along certain directions in the space of its arguments, known as  ambiguity directions. This effect is called the range-delay ambiguity. 

However, a completely deterministic treatment like  that of \cite{ferrara-2017} does not take into account the stochastic effects in scattering, and hence cannot be applied directly to distributed SAR targets~\cite{oliver-98}. A key manifestation of stochasticity in scattering is speckle, which may be thought of as strong and rapid variations of the amplitude and phase of a SAR image while the target parameters of interest remain smooth. Speckle is common in images of most natural and man-made targets when illumination is coherent~\cite{goodman-84}, which is the case for SAR. In the presence of speckle, weak and slow variations along the ambiguity directions in the SAR image can be undetectable, as demonstrated  in~\cite{a85e}. This means that the discrimination between the instantaneous and delayed targets becomes  unreliable, unlike in the deterministic case considered in~\cite{ferrara-2017}.

%
%
A standard approach to problems of this kind is two-fold. To increase the reliability of  classification  one can increase the sample, i.e., the amount of data supplied to the discrimination functional. In~\cite{a85e}, we have demonstrated the advantages of a bigger sample size. To quantify the reliability of  classification outcomes, one needs to employ the confidence levels, which is the primary focus of the current work. Specifically, we refine the discriminating functional for coordinate-delay SAR images, introduce the confidence levels for it, and demonstrate the performance of the discrimination procedure for various system and target parameters. Confidence levels are crucial for applications that rely on the analysis of SAR images because in practice, it is often impossible to obtain additional images of the same target under similar conditions.


Section~\ref{sec:spacetimePSF} presents the coordinate-delay SAR imaging procedure, builds the corresponding imaging operator, and analyzes its properties in terms of the point spread function. In Section~\ref{sec:statmodels}, we introduce models for instantaneous and delayed scatterers and analyze the autocorrelation properties of the resulting coordinate-delay SAR images. Section~\ref{sec:simplifiedStatitstics} presents two models of radar targets to be used in the discrimination problems. 
A binary classification procedure and its extension that uses confidence levels are introduced in Sections~\ref{sec:stRadio_detection} and~\ref{sec:twomatrices}, respectively. In Section~\ref{sec:cdfs}, we analyze the cumulative distribution functions for the values of the discrimination functional and relate them to the quality of discrimination by the original and extended classifier. In the same section, we also introduce the confidence levels and the corresponding threshold values for the discrimination functional in the case of known target contrast. The generalization of confidence levels to all target contrasts is presented in Section~\ref{sec:allcontrasts}. To assess the efficiency of the proposed approach to discrimination between the instantaneous and delayed targets,  we use the Monte-Carlo simulation procedure described in~\cite{a85e}. It lets us build ensembles of sampled SAR images and analyze the statistics of the discriminating functional. 
 Section~\ref{sec:stRadio_results} presents the results of simulation in terms of achievable discrimination quality given a certain confidence level. Section~\ref{sec:stRadio_discussion} discusses possible  future work. Additional bibliography can be found in~\cite{ferrara-2017,a85e}. 

\section{Coordinate-delay SAR image and point spread function}
\label{sec:spacetimePSF}

The goal of SAR imaging is to build an approximate reconstruction of the reflectivity function of the target. In the current formulation, the coordinate-delay reflectivity function $\nu(t_\z,\z)$ defines a relation between the incident and scattered fields denoted by~$u^\incid$ and~$u^\sct$, respectively; this relation is local in space but distributed in delay time~$t_\z$: 
\begin{equation}
\label{eq:linscattering}
\begin{aligned}
	u^\sct(t,\z) 
&=\> 
	\int_0^\infty u^\incid(t-t_\z,\z)\nu(t_\z,\z)\,dt_\z
	.
\end{aligned}
\end{equation}
The lower limit of the integral in~\eqref{eq:linscattering} accounts for the causality of  scattering. 

A synthetic aperture consists of a set of points, $\{\x^n\}$, on the antenna trajectory. The signal emitted by the antenna at each point will be described by $P(t)$, i.e., $u^\incid(t,\x^n) = P(t)$, whereas the scattered field recorded by the antenna is denoted by $u^\sct_{\x^n}(t) \equiv u^\sct(t,\x^n)$; this notation assumes the so-called start-stop approximation, i.e., we ignore the antenna motion during the transmit and receive intervals and between them (see \cite[Chapter~6]{sarbook} for more detail). Additionally, we ignore the propagation attenuation%
\footnote{The propagation attenuation can be factored into the reflectivity function $\nu$, see \cite[Section 2.1.1]{sarbook}. 
}
and assume that the signals emitted from different $\x^n$ don't interfere so that for any $\x=\x^n$ the incident and scattered fields obey the following: 
$$
	u^\incid(t,\z) = P(t-|\x-\z|/c), \quad u^\sct(t,\x) = u^\sct(t-|\x-\z|/c,\z) 
	,
$$
where $c$~is the speed of light. 
Then, the linear model for the field scattered by a distributed non-instantaneous target is as follows: 
\begin{equation}
\label{eq:lippmann}
	u^\sct_{\x^n}(t) 
	= \int_0^\infty dt_\z \int d\z\, \nu(t_\z,\z) P\Big(t-\frac{2R_\z^n}{c}-t_\z\Big) 
	, 
\end{equation}
where $R_\z^n = |\z-\x^n|$. 
In~\eqref{eq:lippmann} and below, the integrals without limits will assume integration over the entire real axis. 
The form~\eqref{eq:lippmann} implies that $\nu$ does not depend on~$\x$, the property called angular coherence, which typically requires that the synthetic aperture is far away from the target and the angle subtended by it and centered at the target is small. 

We build the coordinate-delay image~$I (t_\y,\y)$ by applying the matched filter~$\overline{P(\ldots)}$ to the received signal: 
\begin{equation}
\label{eq:Isum}
	I(t_\y,\y) 
	= \sum_n 
	\int \overline{P\Big(t-\frac{2R^n_\y}{c}-t_\y\Big)} 
	u^\sct_{\x^n}(t) \,dt
	,
\end{equation}
where the overbar denotes the complex conjugate, and $R_\y^n = |\y-\x^n|$. 

Substituting \eqref{eq:lippmann} into~\eqref{eq:Isum}, we obtain the  expression for the imaging operator that relates the SAR image and the reflectivity: 
\begin{equation}
\label{eq:Iconvconv}
	I(t_\y,\y) 
	= \int_0^\infty dt_\z \int \, d\z\, \nu(t_\z,\z) 
	W(t_\y,\y;t_\z,\z)
	,
\end{equation}
with the kernel given by 
\begin{equation}
\label{eq:Wttyz}
	W(t_\y,\y;t_\z,\z) = \frac{N}{\phiT} \int_{-\phiT/2}^{\phiT/2} d\varphi \int dt \,
	 \overline{P\Big(t-\frac{2R^\varphi_\y}{c}-t_\y\Big)}
	P\Big(t-\frac{2R^\varphi_\z}{c}-t_\z\Big)
	.
\end{equation}
In~\eqref{eq:Wttyz}, the summation over~$n$ is replaced with integration over the aperture angle~$\varphi$, $|\varphi| \le \phiT/2 \ll 1$, where $\phiT$ is the angular extent of the synthetic aperture,  with an assumption that the antenna trajectory is an arc of a circle: 
\begin{equation}
\label{eq:ArcOfCircle}
	\x 
	= \x(\varphi) 
	= (x_1(\varphi),x_2(\varphi),x_3(\varphi)) 
	=  \veclist{-L\sin\varphi, -L\cos\varphi,H}
	,
\end{equation}
$L= R\sin\theta$ is the circle radius, $H = R\cos\theta$ is the elevation, $\theta$ is the incidence angle, and $N$~is the total number of the pulse transmit-receive locations. Accordingly, $R^n_\y$ and~$R^n_\z$ in \eqref{eq:lippmann} and~\eqref{eq:Isum} are replaced with $R_\y^\varphi = |\y-\x(\varphi)|$ and~$R_\z^\varphi = |\z-\x(\varphi)|$, respectively. 

Function~$W$ in~\eqref{eq:Iconvconv} is the point spread function (PSF) of the imaging operator in the following sense: $W(t_\y,\y;t_d,\z_d)$ coincides with the image~$I(t_\y,\y)$ due to a space-time point scatterer 
\begin{equation}
\label{eq:spacetimepoint}
	\nu(t_\z,\z) = A \delta(\z-\z_d)\delta(t-t_d)
	,
\end{equation}
where $\delta(\ldots)$ is the Dirac delta function. 
In particular, 
we can see from~\eqref{eq:Wttyz} that $\max |W(t_\y,\y;t_\z,\z)| = \tau N$ is attained when $\y = \z$ and~$t_\y = t_\z$. 

The particular expression for~$W$ requires specification of the pulse shape~$P(t)$ and the distance functions $R_\y^\varphi$ and~$R_\z^\varphi$. We take the standard linear frequency modulated signal, or chirp: 
\begin{equation*}
	P(t)=A(t)e^{-i\omega_0 t},\quad\text{where}\quad A(t)=\chi_\tau(t)e^{-i\alpha t^2}
\end{equation*}
and $\chi_\tau$ is the indicator function:
\begin{equation*}
	\chi_\tau(t)=\begin{cases} 1, & t\in[-\tau/2,\tau/2],\\
	0, & \text{otherwise}.
	\end{cases}
\end{equation*} 
The carrier frequency~$\omega_0$, bandwidth~$B$, duration~$\tau$, and rate~$\alpha$ of the chirp are typically related by 
\begin{equation}
\label{eq:chirpIneqs}
	2\alpha \tau = B \ll \omega_0
\quad \text{and}\quad	
	B\tau \gg 1
\end{equation}
(for simplicity, we only consider $\alpha > 0$). 
For the distances, we take the Taylor expansions of $R_\y^\varphi$ and~$R_\y^\varphi$ up to the second term in~$\varphi$, such that according to~\eqref{eq:ArcOfCircle}, we have 
\begin{equation}
\label{eq:Tphi}
\begin{aligned}
	T^\varphi 
&\bydef\>
	\frac{R_\y^\varphi-R_\z^\varphi}{c} + \frac{t_\y-t_z}{2} 
\\ &\approx\> 
	\Big(\frac{y_2-z_2}{c}\cos\varphi - \frac{y_1-z_1}{c}\sin\varphi\Big)\sin\theta + \frac{t_\y-t_\z}{2} 
\\ &\approx\> 
	\Big[\frac{y_2-z_2}{c}\Big(1-\frac{\varphi^2}{2}\Big) - \frac{y_1-z_1}{c}\varphi\Big]\sin\theta + \frac{t_\y-t_\z}{2} 	
	.
\end{aligned}
\end{equation}
Then, integration in~\eqref{eq:Wttyz} under assumptions~\eqref{eq:chirpIneqs} and~\eqref{eq:Tphi} results in (see details in~\cite{a85e}) 
\begin{equation}
\label{eq:WviaPhi}
	W(t_\y,\y;t_\z,\z) 
	= 
	\tau N 
	e^{-2i\omega_0T^0} \cdot  \Phiarg{(y_1-z_1)}{(y_2-z_2)} \cdot \sinc (B T^0)
	.
\end{equation}
In~\eqref{eq:WviaPhi}, we have introduced 
\begin{equation*}
	\kOtheta \bydef \frac{\omega_0}{c}\sin\theta
\end{equation*} 
and 
\begin{equation*}
	\Phi(v_1,v_2) \bydef \int_{-1/2}^{1/2} e^{2iv_1s}e^{iv_2s^2}ds
	, 
\end{equation*}
while $T^0$ is obtained from~\eqref{eq:Tphi} by setting $\varphi = 0$: 
\begin{equation} 
\label{eq:TO}
	T^0
= 
	\frac{y_2-z_2}{c}\sin\theta  + \frac{t_\y-t_\z}{2} 
	.
\end{equation}

We can see that the list of arguments of~$W$ in~\eqref{eq:WviaPhi} can be contracted to 
\begin{equation}
\label{eq:Wargdiff}
	W(t_\y,\y;t_\z,\z) \equiv W(t_\y - t_\z,\y -\z) 
	.
\end{equation}
The range-delay ambiguity can be observed in the behavior of~$W$ of~\eqref{eq:Wargdiff} along the direction defined by 
\begin{equation}
\label{eq:ambiglines}
	T^0 = \text{const},\quad y_1-z_1=\text{const}
	.
\end{equation}
In the space of arguments of~$W$, 
formula~\eqref{eq:ambiglines} defines a family of straight lines that we call ambiguity lines. 
Consider first the case where $\phiT$ is so small that the quadratic in~$\varphi$ term in~\eqref{eq:Tphi} can be dropped. In this case, the second argument of~$\Phi$ in~\eqref{eq:WviaPhi} is zero, and~\eqref{eq:WviaPhi} reduces to 
\begin{equation}
\label{eq:ambigW}
	\Wlin(t_\y-t_\z,\y-\z) 
	= 
	\tau N 
	e^{-2i\omega_0T^0} \cdot \sinc\big(\kOtheta\phiT(y_1-z_1)\big) \cdot \sinc (B T^0). 
\end{equation}
We can see that $\Wlin$ of~\eqref{eq:ambigW} is constant along the ambiguity lines~\eqref{eq:ambiglines},\eqref{eq:TO}. 
In other words, a delayed scatterer~\eqref{eq:spacetimepoint} is indistinguishable from an  instantaneous scatterer at a point~$\zd'$ on the same ambiguity line, i.e., a scatterer with 
\begin{equation}
\label{eq:ambig_instant} 
	\td' = 0
,\quad
	\zd' = \zd + \frac{ct_\z}{2\sin\theta} \uvector_2
	,
\end{equation}
where $\uvector_2$ is a unit vector in the downrange direction. 
Note that by setting $t_\y=t_\z=0$ in~\eqref{eq:TO} and~\eqref{eq:ambigW} we reduce the latter to the standard SAR formula for the imaging kernel, leading to expressions 
\begin{equation}
\label{eq:DeltaRA}
	\Delta_\az = \pi \frac{1}{\kOtheta\phiT} 
\quad\text{and}\quad  
	\Delta_\rng = \pi \frac{c}{B\sin\theta}
\end{equation}
for the range and azimuthal (i.e., crossrange) resolution, respectively. Obviously, in the coordinate-delay settings \eqref{eq:Iconvconv} and~\eqref{eq:ambigW}, the value of~$\Delta_\rng$ does not completely characterize imaging in range because of the range-delay ambiguity.

Returning to the expression for~$W$ given by \eqref{eq:Tphi}--\eqref{eq:WviaPhi}, we notice that the term~$\Phi(0,\cdot)$ yields unambiguous resolution in the range coordinate. 
Using the explicit form for the marginal function 
\begin{equation}
\label{eq:Phiv2}
	\Phi(0,v_2) = \frac{C(t) + i\sign(v_2)S(t)}{t}
,\quad\text{where}\quad
	t = \sqrt{\frac{|v_2|}{2\pi}}
	,
\end{equation} 
and $C$ and~$S$ are the Fresnel integrals \cite{NIST:DLMF}, we can derive the following asymptotic relations:  
\begin{equation}
\label{eq:asymptPhiv2}
	|\Phi(0,v_2)| \sim 
\begin{cases} 
	(|v_2| / \pi)^{-1/2}
&\quad\text{as}\quad	
	|v_2| \to \infty,
\\
	1 - \frac{1}{360} v_2^2 
&\quad\text{as}\quad	
	v_2 \to 0.
\end{cases}	
\end{equation}	
The width of the main lobe in range due to~\eqref{eq:asymptPhiv2} can be evaluated from $|v_2| \lesssim \sqrt{360} \sim 20$, which corresponds to the distance of 
\begin{equation}
\label{eq:reslnU}
	\Delta_\una \sim \frac{20}{\kOtheta\phiT^2}. 
\end{equation}
At the same time, for the scatterer~\eqref{eq:spacetimepoint} and with~$\zd'$ given by~\eqref{eq:ambig_instant}, we have 
\begin{equation}
\label{eq:ratioOfIs}
	\left|\frac{I(0,\zd')}{I(\td,\zd)}\right| 
	= 
	\left|\frac{W\big({-\td}, c\td/(2\sin\theta) \uvector_2 \big)}{W(0,\0)}\right| 	
	=  
	\left|\Phi\Big(0,\kOtheta \phiT^2 \frac{c\td}{2\sin\theta}\Big)\right|
\end{equation} 

It is possible to interpret formulae \eqref{eq:Phiv2} and~\eqref{eq:ratioOfIs} as follows: 
since $\Phi(0,0) = 1$ and $|\Phi(0,v_2)|<1$ for any~$v_2\ne 0$, we can always discriminate between the pair of delayed and instantaneous scatterers given by \eqref{eq:spacetimepoint} and~\eqref{eq:ambig_instant}, respectively. Namely, any coordinate-delay point $(\td,\zd)$ can be tested for containing a delayed scatterer~\eqref{eq:spacetimepoint} by checking whether this point corresponds to a maximum of~$|I|$ on the ambiguity line~\eqref{eq:ambiglines} passing through it. 
Similarly to the resolution of standard SAR described by \eqref{eq:ambigW} and~\eqref{eq:DeltaRA}, we assume that the location of this maximum can be determined accurate to the width of the main lobe of~$|\Phi(0,v_2)|$  
for $v_2 = \kOtheta \phiT^2 {c\td}/{(2\sin\theta)} = \phiT^2\omega_0 \td/2$, see~\eqref{eq:ratioOfIs}. This yields 
\begin{equation}
\label{eq:td40}
	\phiT^2\omega_0 \td \gtrsim40
\end{equation}
as a detectability condition for a delayed return due to the point scatterer~\eqref{eq:spacetimepoint}. 
Additionally, we introduce the parameter~$\kappa$ to characterize the ratio between two range scales, $\Delta_\rng$ and~$\Delta_\una$, see \eqref{eq:DeltaRA} and~\eqref{eq:reslnU}:  
\begin{equation}
\label{eq:kappadef}
	\kappa \bydef \phiT^2 \frac{\omega_0}{B}
	,
\end{equation}
such that relation~\eqref{eq:td40} can be rewritten as
\begin{equation}
\label{eq:td40p}
\tagprime{eq:td40}
	\kappa \frac{B\td}{2} \gtrsim 20. 
\end{equation}

On the way to more realistic setups, we are going to include a homogeneous background and consider a certain range of delay times rather than a fixed delay~$\td$ as in~\eqref{eq:spacetimepoint}. These changes introduce a new effect, called speckle, into the consideration. In the following sections, we will formulate the metrics of detectability of a delayed return in the presence of speckle. 


\section{Statistical coordinate-delay models of distributed radar scatterers}
\label{sec:statmodels}

The deterministic description of SAR imaging presented in Section~\ref{sec:spacetimePSF} is appropriate for the space-time point scatterers given by~\eqref{eq:spacetimepoint}. However, once the support of~$\nu$ becomes nonsingular in any argument, a statistical description appears to be a proper way of describing interference of multiple scatterers within one resolution cell (see details in \cite{a85e,oliver-98,goodman-84}). 

In particular, a homogeneous instantaneous reflectivity, or the background, is modelled by  
\begin{equation}
\label{eq:modelz}
	\text{\underline{background}:}\quad 
	\nuz(t_\z,\z) = 
	\delta(t_\z)\muz(\z)
	,
\end{equation}
where $\muz(\z)$ is a two-dimensional circular Gaussian white random field: 
\begin{equation}
\label{eq:muzmuz}
	\avg{\muz(\z)} = 0
,\quad 
	\avg{\muz(\z)\muz(\z')} = 0
,\quad 
	\big\langle\overline{\muz(\z)} \muz(\z') \big\rangle 
=
	\sigmasqz \delta (\z - \z')
	.
\end{equation}
In~\eqref{eq:muzmuz}, 
$\langle\ldots\rangle$ denotes statistical averaging 
and~$\sigmasqz$ is a positive deterministic constant describing the statistically averaged reflectivity~\cite{oliver-98}. When a random reflectivity function~\eqref{eq:modelz}--\eqref{eq:muzmuz} is substituted into the imaging operator~\eqref{eq:Iconvconv}, the resulting image~$I_\text{b}(t_\y,\y)$ is a stationary circular Gaussian random field, as confirmed in numerous experiments~\cite{ulaby-89,oliver-98,goodman-84}. For the lower moments of such field, we have 
\begin{equation}
\label{eq:specklemoments}
	\avg{I_\text{b}(t_\y,\y)} = 0
 ,\quad 
	\avg{I_\text{b}^2(t_\y,\y)} 
	= 0 
,\quad 
	\Var\big(|I_\text{b}(t_\y,\y)|^2\big) = \avg{|I_\text{b}(t_\y,\y)|^2}^2
	.
\end{equation}
The suggested physical model behind this behavior is that each resolution element is populated by a large number of  uncorrelated point scatterers such that each image pixel is a result of interference of individual returns with homogeneously distributed phases \cite{oliver-98,goodman-84}. 

In addition to the spatially homogeneous model \eqref{eq:modelz}--\eqref{eq:muzmuz}, we introduce two models of inhomogeneous scatterers. First, we want to model a scatterer that is small in size and exhibiting a certain range of delays. Such scatterer, later called  t-scatterer, may be representative of an opening into some cavity, e.g., a manhole or window in a wall. Multi-path reflections or structural dispersion inside this enclosed space will result in a range of response delays; if there are many such paths (or many electromagnetic cavity eigenmodes), then the responses with different delay times can be considered essentially uncorrelated, similarly to the scatterers at different locations in the model \eqref{eq:modelz}--\eqref{eq:muzmuz}. Hence, we modify the latter model as follows: 
\begin{equation}
\label{eq:modeltz}
	\text{\underline{t-scatterer}:}\quad 
	\nut(t_\z,\z) \equiv \nut(t_\z,\z; \zd) =  
	\mut(t_\z)\delta(\z-\zd)
	,
\end{equation}
where $\mut(t)$ is a one-dimensional circular Gaussian white random process with the following properties:
\begin{equation}
\label{eq:mutmut}
	\avg{\mut(t)} = 0
,\quad
	\big\langle \mut(t) \mut(t') \big\rangle = 0
,\quad
	\big\langle\overline{\mut(t)} \mut(t') \big\rangle 
	= 
	\sigmasqt F_t(Bt/2) \delta (t-t') 
	. 
\end{equation} 
In~\eqref{eq:modeltz}, $\zd$ is a location of the t-scatterer (cf.~\eqref{eq:spacetimepoint}), whereas the product of a positive constant $\sigmasqt$ and a non-negative dimensionless function  $F_t(Bt/2)$ in~\eqref{eq:mutmut} describes averaged reflectivity as a function of delay time. We require certain properties of~$F_t(\zeta)$, in particular, that $F_t(\zeta) = 0$ for $\zeta < 0$ from the causality considerations (cf.~\eqref{eq:linscattering}) and the integrability to satisfy a sufficient condition for the existence of the process~\eqref{eq:mutmut} \cite{allen-97}. 
Hence, $\mut$ is a nonstationary circular Gaussian white noise, and the moments of the image~$I_t(t_\y,\y)$ due to such reflectivity function will obey the same relations as~$I_\text{b}$ in~\eqref{eq:specklemoments}. 

Similarly to~\eqref{eq:modeltz}--\eqref{eq:mutmut}, we define an instantaneous inhomogeneous scatterer with the support on a straight line drawn in the range direction: 
\begin{equation}
\label{eq:modelsz}
	\text{\underline{s-scatterer}:}\quad
	\nus(t_\z,\z) \equiv \nus(t_\z,\z; \zd) = 
	\delta(t_\z) \delta(z_1-\zdO) \mus(z_2-\zdT) 
	,
\end{equation}
where $\mus(s)$ is another one-dimensional inhomogeneous circular Gaussian white random process: 
\begin{equation}
\label{eq:musmus}
	\avg{\mus(s)} = 0
,\quad
	\big\langle \mus(s) \mus(s') \big\rangle = 0
,\quad
	\big\langle\overline{\mus(s)} \mus(s') \big\rangle
=
	\sigmasqs F_s(B\kOtheta s / \omega_0) \delta (s-s') 	
	.
\end{equation} 
The support of the scatterer in~\eqref{eq:modelsz}--\eqref{eq:musmus} is related to that in~\eqref{eq:modeltz}--\eqref{eq:mutmut} through the ambiguity relation illustrated by~\eqref{eq:ambig_instant}. 
This means that images due to these two scatterers, $I_t(t_\y, \y)$ and $I_s(t_\y,\y)$, may resemble each other, and the problem of detection of a delayed scatterer may be formulated as a problem of discrimination between these two cases. In this  context, it makes sense to assume that $F_s(\zeta) = F_t(\zeta)$. 

Introduce the dimensionless coordinates~$\etad$, $\zetad$, and~$\psid$ with the origin at certain~$(\td,\zd)$ as follows: 
\begin{equation}
\label{eq:etazetapsid}
\begin{aligned}
	\etad 
&=\>
	\kOtheta \phiT(y_1-\zdO)
	,
\\
	\zetad
&=\>
	\frac{B}{\omega_0}\kOtheta (y_2 - \zdT)+ B\frac{t_\y-\td}{2}
	,
\\
	\psid
&=\>
	\frac{B}{\omega_0}\kOtheta (y_2 - \zdT) - B\frac{t_\y-\td}{2}	
	,
\end{aligned}
\end{equation}
This way, the coordinate~$\psid$ is aligned with the ambiguity line~\eqref{eq:ambiglines}. We take $\td=0$ for simplicity and present the second order statistics 
of images due to the scatterer models (\ref{eq:modelz}), (\ref{eq:modeltz}), and (\ref{eq:modelsz}), for $\etad=\etad'=0$ and $\zetad=\zetad'$: 
\newcommand{\allalphas}{\alpha}
\begin{equation}
\label{eq:psidpsidp}
	\avg{I_\allalphas (\zetad,\psid) \roverline{I_\allalphas (\zetad,\psid')}} 
	= \sigma^2_\allalphas K_\allalphas H_\allalphas(\zetad,\psid,\psid')
	.
\end{equation}
In~\eqref{eq:psidpsidp}, the following notations are used. 

\begin{itemize}

\item
For $\alpha \in \{\text{b},s,t \}$ denoting the scatterer types given by \eqref{eq:modelz}, \eqref{eq:modelsz}, and~\eqref{eq:modeltz}, respectively, formula~\eqref{eq:psidpsidp} is obtained directly by substituting the expressions  for~$\nu_\alpha(t_\z,\z)$ into~\eqref{eq:Iconvconv} and~\eqref{eq:WviaPhi}. We can choose $H_\alpha$ to be dimensionless with $\sup |H_\alpha| \sim 1$ as follows: 
\begin{equation}
\label{eq:HbHtHs} 
\begin{aligned}
	H_\text{b}(\zeta,\psid,\psid')
=&\>	
	\Phi\Big(0,\kappa\frac{\psid-\psid'}{2}\Big)
	,
\\
	H_t(\zetad,\psid,\psid')
=&\>	
	\Phi\Big(0,\kappa\frac{\zetad+\psid}{2}\Big)
	\roverline{
	\Phi\Big(0,\kappa\frac{\zetad+\psid'}{2}\Big)
	}
\\ \cdot & \>
	\frac{1}{\pi}
	\int_0^\infty F_t^2(\zeta) \sinc^2(\zetad-\zeta)\,d\zeta
	,
\\
	H_s(\zetad,\psid,\psid')
=&\>	
	\frac{1}{\pi}
	\int_0^\infty F_s^2(\zeta) \sinc^2(\zetad-\zeta)
\\ & \cdot 	
	\Phi\Big(0,\kappa\frac{\zetad+\psid}{2} - \kappa\zeta\Big)
	\roverline{
	\Phi\Big(0,\kappa\frac{\zetad+\psid'}{2} - \kappa\zeta\Big)
	}
	\,d\zeta
	,
\end{aligned}
\end{equation}
such that the normalizing coefficients are given by
$$
	K_\text{b}  = 
	N^2\tau^2 \frac{\omega_0}{B\kOtheta} \frac{1}{\kOtheta\phiT} \cdot 
	\pi^2
,\quad
	K_t = N^2\tau^2\frac{2}{B}\pi
,\quad
	K_s = N^2\tau^2\frac{\omega_0}{B\kOtheta} \pi 
	.
$$

\item 
We introduce the noise term~$I_\text{n}$ to represent the  receiver noise and processing errors. We do so by allowing $\alpha = \text{n}$ in~\eqref{eq:psidpsidp} and formally setting  
\begin{equation}
\label{eq:Hn}
	H_\text{n}(\zetad,\psid,\psid') = 0 
\quad\text{if $\psid \ne \psid'$, \quad and}	 \quad
	H_\text{n}(\zetad,\psid,\psid') = 1 
\quad\text{if $\psid = \psid'$}	
	,
\end{equation}
so that all noise terms are uncorrelated. At the same time, the value of~$\sigma^2_\text{n}K_\text{n}$ is a separate problem parameter as specified in~\eqref{eq:contrastdef} below.

\end{itemize} 

In Fig.~\ref{fig:IsIt}, we plot expectations for image intensities, $\avg{|I_t|^2}$ and~$\avg{|I_s|^2}$, in the plane $y_1 - \zdO = 0$ (cf.~\eqref{eq:ambiglines}) for functions 
\begin{equation}
\label{eq:finitesupportF}
	F_t=F_s = \text{\bf 1}_{[0,\zetamax]}, 
\end{equation}
with different values of $\zetamax$. This is done by setting  $\psid = \psid'$ in \eqref{eq:psidpsidp} and~\eqref{eq:HbHtHs}. As $\zetamax$ increases, the parallelogram-shaped level lines of~$|H_t|$ and~$|H_s|$ stretch in vertical and horizontal directions, respectively, which is in agreement with the analysis made in~\cite{a85e}. The size of the parallelograms in the direction along the ambiguity lines is determined by the width of the main lobe of~$\Phi(0,\kappa\zeta)$, see~\eqref{eq:asymptPhiv2}. 
It can be argued that the shapes in one column of Fig.~\ref{fig:IsIt} differ substantially from the respective shapes in the other column if this width is smaller than the support of $F_t$ and~$F_s$, i.e., 
\begin{equation}
\label{eq:kz20}
	\kappa\zetamax \gtrsim 20
	,  
\end{equation} 
cf.~\eqref{eq:td40p}. 
Hence, the ``difference'' between the plots of $\avg{|I_t|^2}$ and~$\avg{|I_s|^2}$ can be increased by increasing either~$\zetamax$ or~$\kappa$ (or both). 

It appears quite feasible to apply traditional image processing techniques, such as edge detection and segmentation \cite{canny-86,basu-02,ziou-98,marr-80}, to the shapes in Fig.~\ref{fig:IsIt} in order to determine the type of the scatterer and its parameters, such as~$\zetamax$. 
However, the intensities of  actual images look dramatically different from their statistical averages because of the speckle, and while in practice there is typically only a single image acquisition of the scene of interest, computation of statistical averages from the empirical data is ruled out. 
For images with speckle, such as simulated in Fig.~\ref{fig:noisyIsIt}, the mere detection (let alone classification) of the target in certain cases looks problematic. The goal of the next sections is to quantify our ability to distinguish between the t-scatterer and s-scatterer as defined in \eqref{eq:modeltz}--\eqref{eq:musmus} in the presence of speckle.

\begin{figure}[ht!]
\begin{center}
	\includegraphics[width=5.4in]{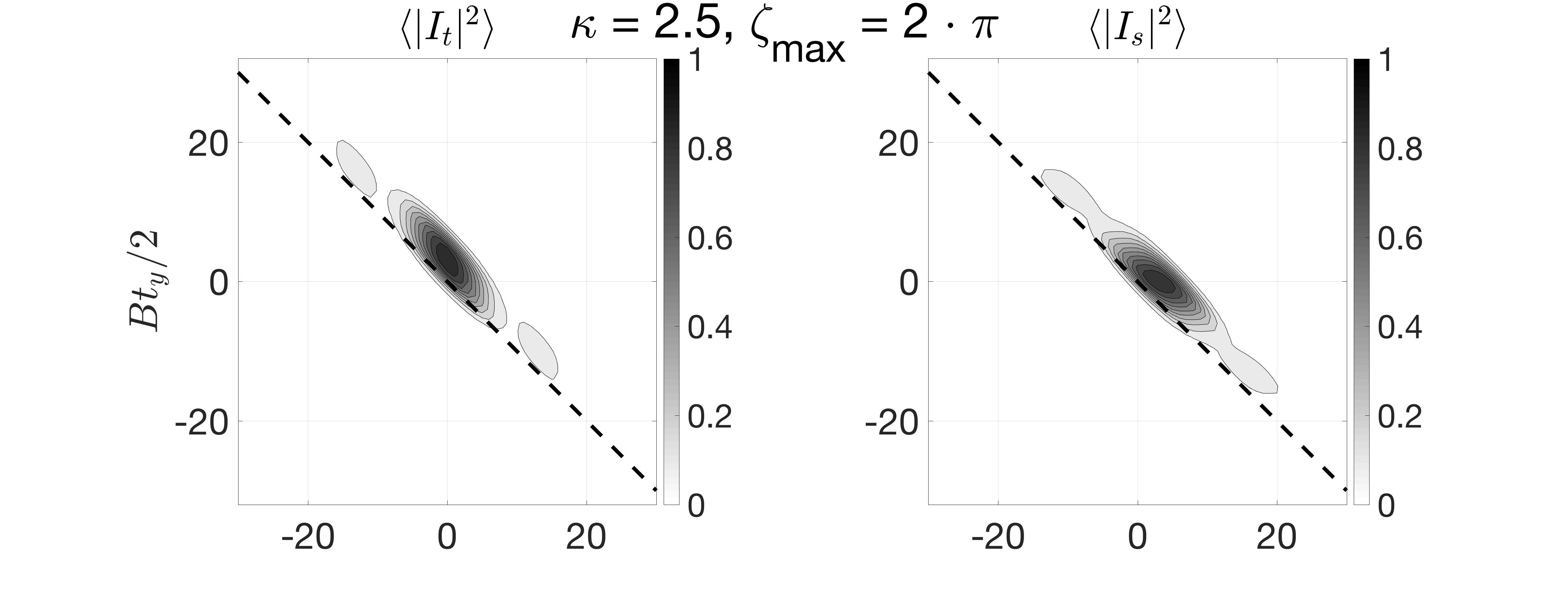}
	\includegraphics[width=5.4in]{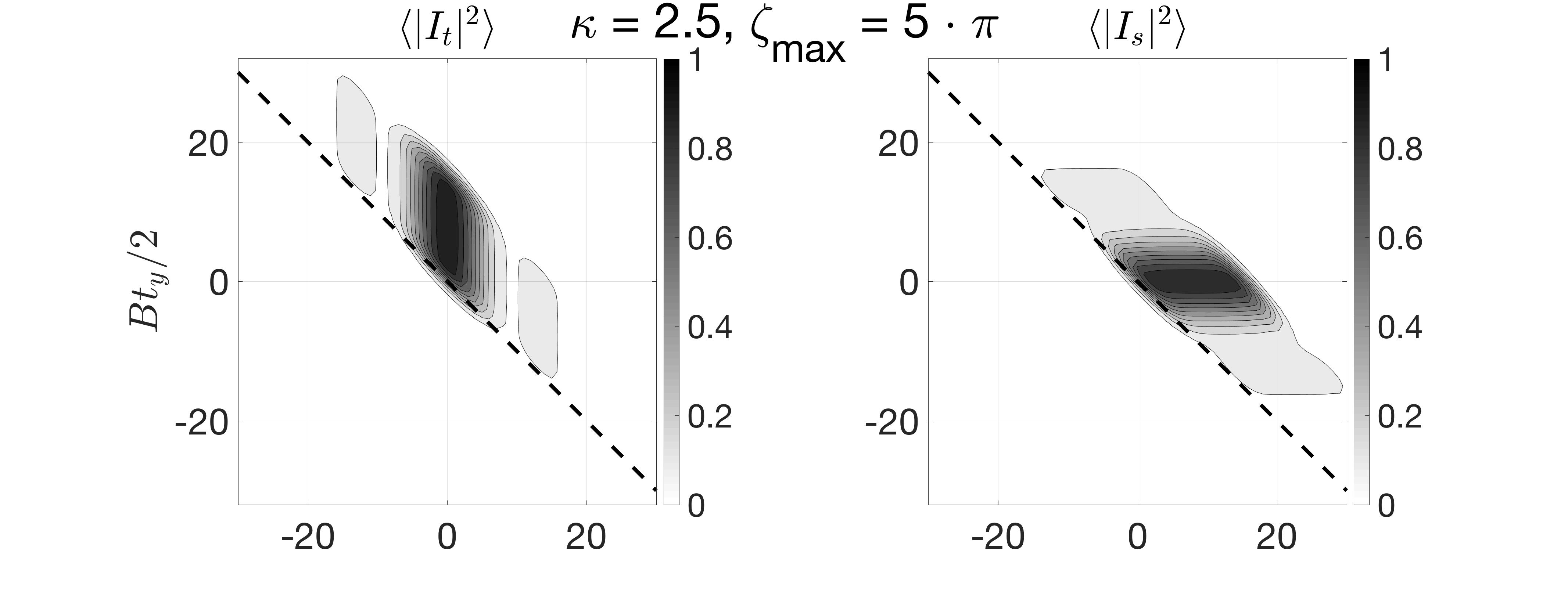}
	\includegraphics[width=5.4in]{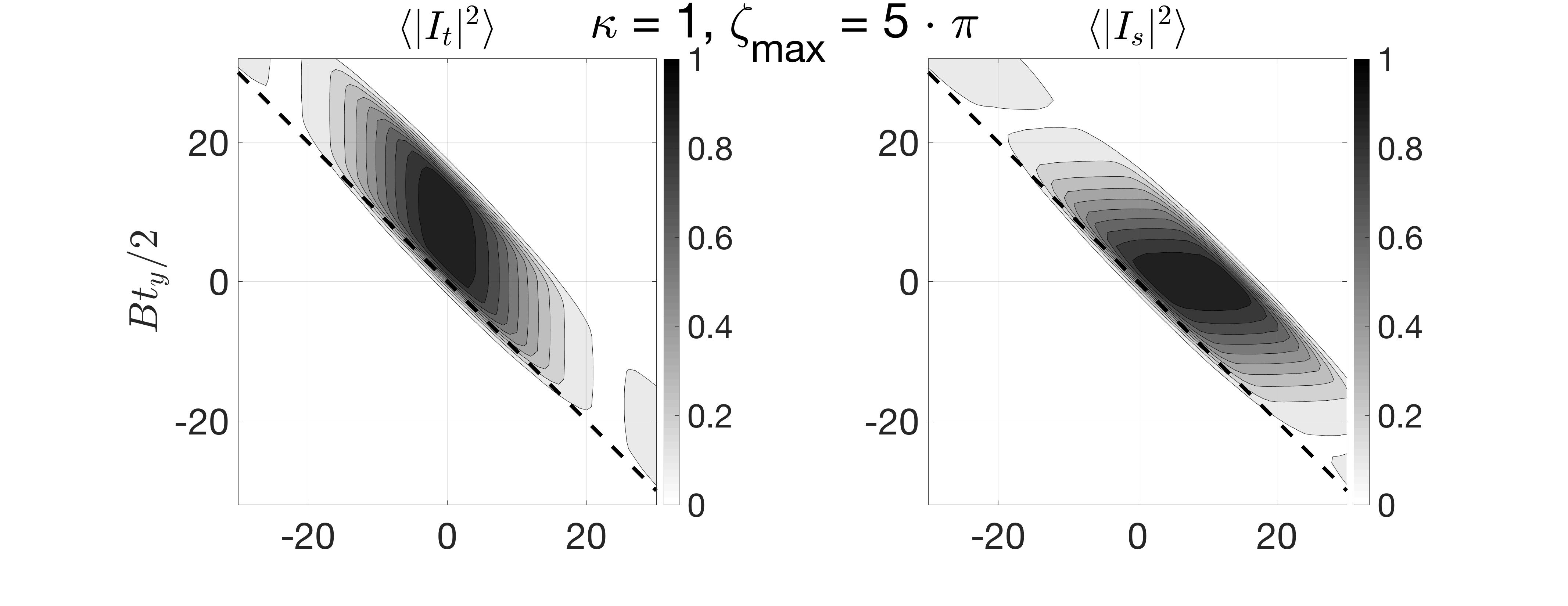}
\vspace{-0.2cm}
\end{center}
\caption{
Plots of $\avg{|I_t|^2}$ and~$\avg{|I_s|^2}$ for different values of $\zetamax$ and~$\kappa$, 
see \eqref{eq:finitesupportF} and~\eqref{eq:kappadef}. The dashed lines passing through the origin indicate the ambiguity direction, see~\eqref{eq:TO},\eqref{eq:ambiglines}. For the middle row of plots, the condition $\kappa\zetamax \gtrsim 20$ (see~\eqref{eq:kz20}) is satisfied, and the difference in the orientation of the parallelogram-shaped level lines is more apparent than for the top and bottom rows.
}
\label{fig:IsIt}
\end{figure}

\section{Two models for  coordinate-delay SAR images}
\label{sec:simplifiedStatitstics}

Using the scatterer types described in Section~\ref{sec:statmodels}, we are going to build models of radar targets to be used in discrimination problems. For simplicity, we assume that there are two possible configurations of scatterers in the target: 
\begin{subequations}
\label{eq:6nu}
\begin{equation}
\label{eq:6nus}
	\nu(t_\z,\z) = 
	\nu_\text{s-model}(t_\z,\z;\zd) = \nuz(t_\z,\z)+\nus(t_\z,\z;\zd)
\end{equation}
and  
\begin{equation}
\label{eq:6nut}
	\nu(t_\z,\z) = 
	\nu_\text{t-model}(t_\z,\z;\zd) = \nuz(t_\z,\z)+\nut(t_\z,\z;\zd)
	,
\end{equation}
\end{subequations}
where $\nuz$, $\nut$, and $\nus$ are defined in~\eqref{eq:modelz}, \eqref{eq:modeltz}, and~\eqref{eq:modelsz}, respectively, and~$\zd$ in \eqref{eq:6nus} and~\eqref{eq:6nut} is the same. The names ``s-model'' and ``t-model'' are intended to match the terms ``s-scatterer'' and ``t-scatterer'' introduced in Section~\ref{sec:statmodels}, see~\eqref{eq:modeltz} and~\eqref{eq:modelsz}. The coordinate-delay SAR images resulting from substitution of~\eqref{eq:6nu} into~\eqref{eq:Iconvconv}  are then given by either  
\begin{subequations}
\label{eq:16Ity}
\begin{equation}
\label{eq:16Itys}
	I_\text{s-model}(t_\y,\y;\zd) = I_\text{b}(t_\y,\y) + I_\text{n}(t_\y,\y)  + I_s(t_\y,\y;\zd) 
	\phantom{.}
\end{equation}
or 
\begin{equation}
\label{eq:16Ityt}
	I_\text{t-model}(t_\y,\y;\zd) = I_\text{b}(t_\y,\y)  + I_\text{n}(t_\y,\y)+ I_t(t_\y,\y;\zd)
	, 
\end{equation}
\end{subequations}
where the terms~$I_\text{n}$ are described by~\eqref{eq:psidpsidp} with~\eqref{eq:Hn}. 

For  delta-correlated scatterers as in Section~\ref{sec:statmodels}, the correlation of the image described   by~\eqref{eq:Iconvconv} is determined by the properties of the imaging kernel~$W$. In particular, the correlation of the image rapidly decreases across the ambiguity lines~\eqref{eq:ambiglines} owing to the sinc term in~\eqref{eq:WviaPhi}. This allows us to simplify the presentation of the correlation function of images given by~\eqref{eq:16Ity} by specifying a discrete set of ambiguity lines with large enough spacing between them such that the image values at different lines can be considered uncorrelated. 

We assume that the terms $I_\text{b}$, $I_\text{n}$, $I_s$, and~$I_t$ in each line of~\eqref{eq:16Ity} are independent;  hence,  the moments of the total image are sums of the moments of the corresponding components. In turn, due to the Gaussianity, the moments of each component can be calculated using formula~\eqref{eq:psidpsidp}. Consequently, formulae \eqref{eq:psidpsidp}--\eqref{eq:Hn} provide a complete description of the statistics of the total image for the arguments corresponding to one and the same ambiguity line. 

\begin{figure}[ht!]
\begin{center}
	\includegraphics[width=5.5in]{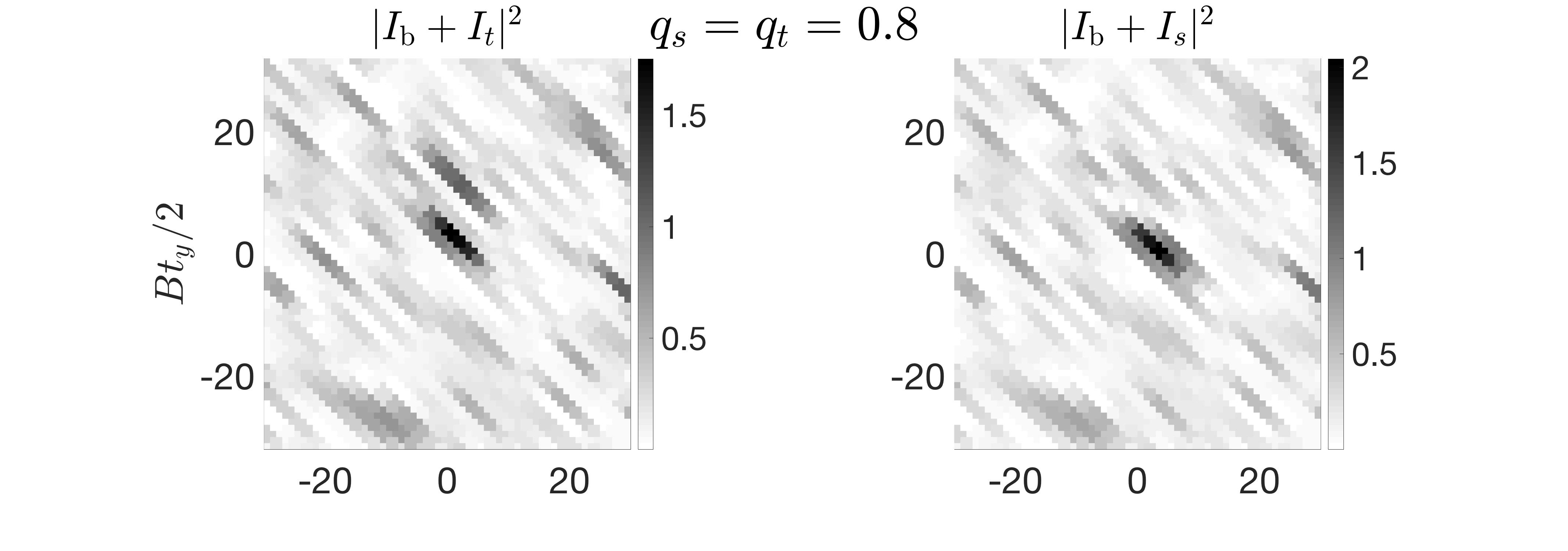}
\vspace{-0.2cm}
	\includegraphics[width=5.5in]{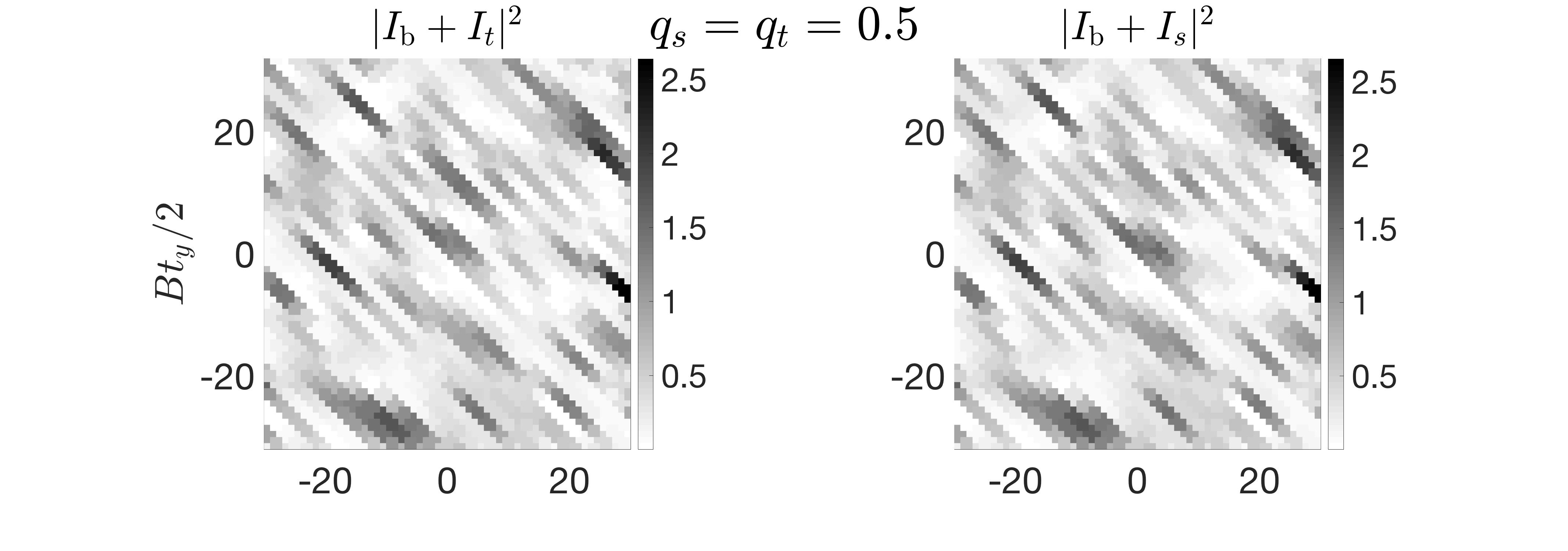}
\vspace{-0.2cm}
	\includegraphics[width=5.5in]{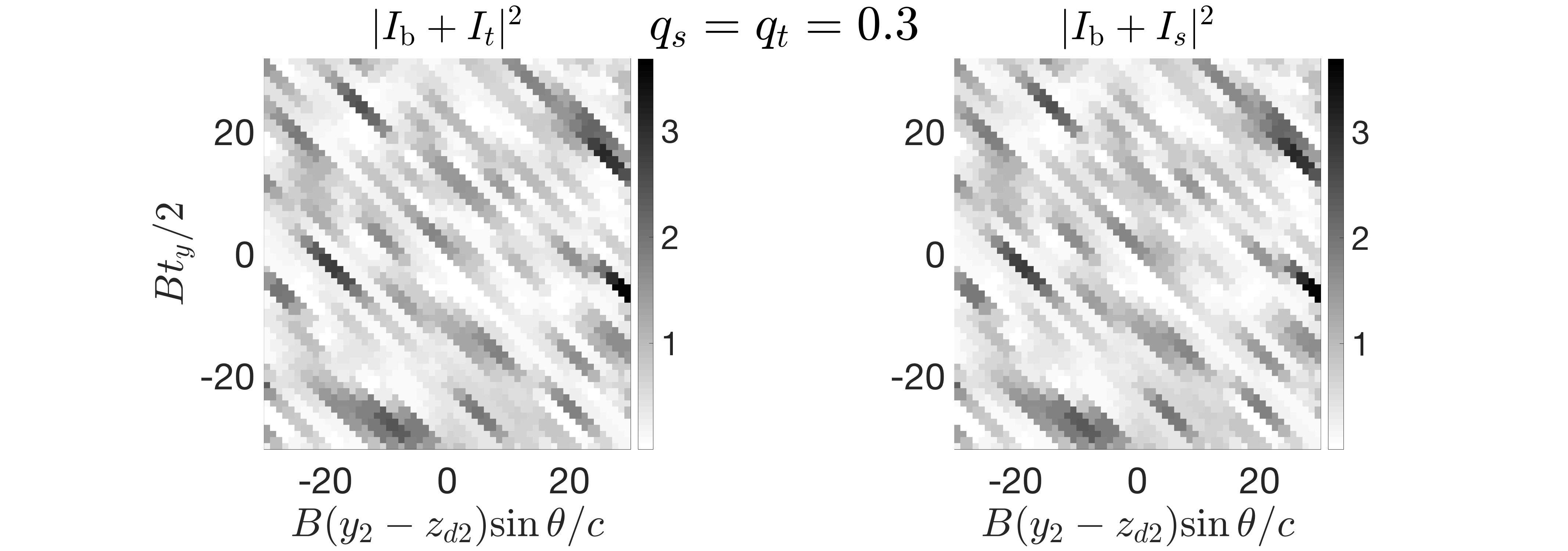}
\vspace{-0.6cm}
\end{center}
\caption{
Simulated coordinate-delay SAR images with different contrasts. We use $\kappa = 2.5$ and~$\zetamax = 5\pi$, which corresponds to the middle row in Fig.~\ref{fig:IsIt}. 
To build each image, we sample $\zetad$ and~$\psid$ with a step of~$\pi$ each, and for every~$\zetad$ (i.e., each ambiguity line) generate the multivariate circular Gaussian image components according to~\eqref{eq:psidpsidp} with the weights $ \sigma^2_\allalphas K_\allalphas$ calculated via~\eqref{eq:contrastdef}. The total images are then computed according to the appropriate expression in~\eqref{eq:16Ity}. Simulation of multivariate normal random variables is performed using the $\text{MATLAB}^\copyright$  function {\tt mvnpdf}. 
}
\label{fig:noisyIsIt}
\end{figure}

Fig.~\ref{fig:noisyIsIt} shows examples of simulated coordinate-delay SAR images due to the targets~\eqref{eq:6nus} (right panels) and~\eqref{eq:6nut} (left panels) with $F_s$ and~$F_t$ given by~\eqref{eq:finitesupportF}. 
The relative scatterer intensities, or contrasts, are defined as follows: 
\begin{equation}
\label{eq:contrastdef}
	p_\text{n} = \frac{\sigmasqn K_\text{n}}{\sigmasqz K_\text{b}}
,\quad
	q_s 
		= \frac{\sigmasqs K_s}{\sigmasqs K_s + \sigmasqz K_\text{b} + \sigmasqn K_\text{n}}
,\quad 
	q_t
             	= \frac{\sigmasqt K_t}{\sigmasqt K_t + \sigmasqz K_\text{b} + \sigmasqn K_\text{n}}
	.
\end{equation}	
Note that for Fig.~\ref{fig:noisyIsIt}, we have chosen $p_\text{n} = 0$, i.e., no noise component, while the target contrasts~$q_s=q_t=q$ take three different values. It can be seen that the visible shape features distinguishing the two types of scatterers in Fig.~\ref{fig:IsIt} appear much less prominent even for a high contrast of $q  = 0.8$ in the top row of Fig.~\ref{fig:noisyIsIt}, and practically disappear for the lower  contrasts.  We will see that the value of~$q$ is very important for the effectiveness of the discrimination algorithms described in Sections~\ref{sec:stRadio_detection} and~\ref{sec:twomatrices}. The effect of the value of~$p_\text{n}$ has not been as prominent, and we always set it to $p_\text{n}=0.1$.

\section{Detection of the delayed response}
\label{sec:stRadio_detection}

In this work, we reduce the problem of detection of the delayed response to discrimination between the scenario~\eqref{eq:6nus} involving only instantaneous scatterers and~\eqref{eq:6nut} that includes a component~$\nut$ with a scattering delay. 

Assume that we have some observation data $u^\sct_{\x^n}(t)$, see~\eqref{eq:lippmann}. Using~\eqref{eq:Isum}, we can build a coordinate-delay image~$I(t_\y,\y)$. We also assume that we can identify candidates for~$\zd$ as locations of a sharp increase of the image intensity along the range direction at~$t_\y = 0$, see Figs.~\ref{fig:IsIt},\ref{fig:noisyIsIt}.%
\footnote{This can be done using one of the standard edge detection methods \cite{canny-86,basu-02,ziou-98,marr-80}.} 
After that, the neighborhood of each  candidate location $\zd$ goes through the discrimination procedure described below. This procedure attributes the apparent inhomogeneity at~$\zd$ to one of the two classes in~\eqref{eq:6nu}. 

Let $\{\zeta_m\}$ be a discrete set of values of~$\zetad$, see~\eqref{eq:etazetapsid}, for some~$\zd$.  
In particular, we define this set according to 
\begin{equation}
\label{eq:zetaym}
	\zeta_m = \pi m
,\quad\text{where}\quad	
	m\in\mathbb{N}
,\quad	
	\zetamin \le \zeta_m \le \zetamax
	.
\end{equation}	
In~\eqref{eq:zetaym}, we have  introduced another parameter, $\zetamin =3 \pi$, to cut off the transitional effects due to the behavior  of~$F_s(\zeta)$ and~$F_t(\zeta)$ given by~\eqref{eq:finitesupportF} in the vicinity of~$\zeta = 0$. Each value of~$\zeta_m$ defines an ambiguity line passing through a neighborhood of~$(0,\zd)$, and, according to the discussion in Section~\ref{sec:simplifiedStatitstics}, the spacing of~$\pi$ between the adjacent values of~$\zeta_m$ allows us to consider the values on different ambiguity lines uncorrelated. 

For each~$m$, we choose $N_m$~values of~$\psi_{mj}$, $1 \le j \le N_m$; these values will play the role of~$\psid$ and~$\psid'$ for a given~$\zetad = \zeta_m$ in~\eqref{eq:psidpsidp}. Then, $I_{mj}$ will denote the coordinate-delay SAR image sampled in a neighborhood of~$(0,\zd)$. 
We can represent the second order statistics for the expressions in~\eqref{eq:16Ity} with the help of~\eqref{eq:psidpsidp} in the following form: 
\begin{equation}
\label{eq:sumcalA}
\begin{aligned}
	\avg{I_{mj,\,\text{s-model} } \; \roverline{I_{mj',\,\text{s-model} }} } 
=&\> 
	 \sum_{\alpha \in {\cal S}}	
	  \sigma^2_\allalphas K_\allalphas H_\allalphas(\zeta_m,\psi_{mj},\psi_{mj'})
,\quad  
	1 \le j,\!j' \le N_m,
\\
	\avg{I_{mj,\,\text{t-model} } \; \roverline{I_{mj',\,\text{t-model} }} } 
=&\> 
	 \sum_{\alpha \in {\cal T}}	
	  \sigma^2_\allalphas K_\allalphas H_\allalphas(\zeta_m,\psi_{mj},\psi_{mj'})
,\quad  
	1 \le j,\!j' \le N_m,
\end{aligned}
\end{equation}
where 
\begin{equation}
\label{eq:calA}
	{\cal S} = \{\text{b},\text{n},s\} 
,\quad 
	{\cal T} = \{\text{b},\text{n},t\}.
\end{equation}
Remember that the statistical averages in~\eqref{eq:sumcalA} are unavailable in a practical setting. 
Instead, we will use the actual data~$I_{mj} \roverline{I_{mj'}} $ and  for each line of~\eqref{eq:sumcalA} build an objective function for optimization with the unknown scatterer intensities as the design variables. 
 For each of the two scenarios in~\eqref{eq:16Ity}, i.e., for ${\cal S}$ and ${\cal T}$ in~\eqref{eq:calA},  our discrimination algorithm will seek a set of values for unknowns~$\sigmasqalpha$ that maximizes the probability density of the image with the statistics described by~\eqref{eq:sumcalA}. Then, we will choose the model that yields the larger of the two maxima. 
Essentially, this is a maximum likelihood (ML) based procedure~\cite{Mendenhall-73,oliver-98}. 

The probability density of the sampled image~$\{I_{mj}\}$ for either of the two models 
is calculated as follows. For each~$m$ we create a real-valued vector $\mathbf{r}_m$ of dimension~$2N_m$: 
\newcommand{\TT}{{\text{{\bf T}}}}
\begin{equation}
\label{eq:vecrmdef}
	\mathbf{r}_m = \veclist{\myRe\,I_{m1}, \myIm\,I_{m1},
				       \myRe\,I_{m2}, \myIm\,I_{m2}, \dots
				       \myRe\,I_{mN_m}, \myIm\,I_{mN_m}  
	}^\TT
	.
\end{equation}
Then, due to the circular Gaussianity and independence of all~$I_\alpha$, formula~\eqref{eq:sumcalA} can be recast as 
\newcommand{\Mm}{\mathbf{M}^{(m)}}
\begin{equation}
\label{eq:rmrmT}
	\avg{\mathbf{r}_m\mathbf{r}_m^\TT} 
	\bydef \Mm = 
  \begin{pmatrix} 
    \Mm_{11} & \Mm_{12} & \dots  & \Mm_{1N_m} \\
    \Mm_{21} &  \Mm_{22} & \dots  & \Mm_{2N_m} \\   
    \vdots       &                                    & \ddots & \\
    \Mm_{N_m1} & \Mm_{N_m2}          & \dots  & \Mm_{N_mN_m} 
    \end{pmatrix}
    , 
\end{equation}
where each individual $2\times 2$ block on the right hand side is given by 
\begin{equation}
\label{eq:Mjjp}
	\Mm_{jj'} = \frac{1}{2}
	\sum_{\alpha \in {\cal A}}	
	 \sigma^2_\allalphas K_\allalphas \cdot
	 \begin{pmatrix}
	 	  \myRe\, H_\alpha(\zeta_m, \psi_{mj}, \psi_{mj'}) 
		&
	 	- \myIm\, H_\alpha(\zeta_m, \psi_{mj}, \psi_{mj'})	
	\\
	 	 \myIm\, H_\alpha(\zeta_m, \psi_{mj}, \psi_{mj'})
		&
	 	\phantom{-}\myRe\, H_\alpha(\zeta_m, \psi_{mj}, \psi_{mj'} )			
	 \end{pmatrix}.
\end{equation}
Choosing ${\cal A} = {\cal S}$ or ${\cal A} = {\cal T}$ in~\eqref{eq:Mjjp}, we obtain two expressions for the matrices~$\Mm$ in~\eqref{eq:rmrmT}, henceforth called~$\Mm_\text{s-model}$ and~$\Mm_\text{t-model}$. These matrices give rise to two multivariate Gaussian distribution functions: 
\begin{equation}
\label{eq:pmGaussian}
\begin{aligned}
	p_\text{s-model}(\mathbf{r}_m)
=&\> 
	\frac{1}{ (2\pi)^{N_m} \, \sqrt{ \det \Mm_\text{s-model} }}
	\exp\Big({-\frac{1}{2}\mathbf{r}_m^\TT(\Mm_\text{s-model})^{-1}\mathbf{r}_m}\Big),
\\ 
	p_\text{t-model}(\mathbf{r}_m)
=&\> 
	\frac{1}{ (2\pi)^{N_m} \, \sqrt{ \det \Mm_\text{t-model} }}
	\exp\Big({-\frac{1}{2}\mathbf{r}_m^\TT(\Mm_\text{t-model})^{-1}\mathbf{r}_m}\Big)
	.
\end{aligned}
\end{equation}
Then, we extend formulae~\eqref{eq:pmGaussian} by including the data from multiple ambiguity lines given by a set of~$\zeta_m$. In a simplified treatment suggested in Section~\ref{sec:simplifiedStatitstics}, we consider the data for different~$\zeta_m$ uncorrelated. 
Then, for the full dataset vector~$\mathbf{R}$ that combines all $\mathbf{r}_m$-vectors~\eqref{eq:vecrmdef}: 
\begin{subequations}
\label{eq:datasetRm}
\begin{gather}
\label{eq:datasetRm_a}
	\mathbf{R} = \veclist{\mathbf{r}^\TT_1, \mathbf{r}^\TT_2, \ldots, \mathbf{r}_m^\TT, \ldots}^\TT,\\
\intertext{we have}
\label{eq:datasetRm_b}
	p(\mathbf{R}) = \prod_m p(\mathbf{r}_m).
\end{gather}
\end{subequations}

The actual dataset vector~$\mathbf{Q}$ representing a sampled image has the same structure as~$\mathbf{R}$ of~\eqref{eq:datasetRm_a}:
\begin{equation}
\label{eq:datasetQ}
	\mathbf{Q} = \veclist{\mathbf{q}^\TT_1, \mathbf{q}^\TT_2, \ldots, \mathbf{q}_m^\TT, \ldots}^\TT,
\end{equation}
where each vector~$ \mathbf{q}_m$ corresponds to image values taken at a certain ambiguity line. We will then consider (cf.~\eqref{eq:pmGaussian}, \eqref{eq:datasetRm_b}) 
\begin{equation}
\label{eq:pQ}	
\begin{aligned}
	p_\text{s-model}(\mathbf{Q}) = &\> \prod_m 
	\frac{1}{ (2\pi)^{N_m} \, \sqrt{ \det \Mm_\text{s-model} }}
	\exp\Big({-\frac{1}{2}\mathbf{q}_m^\TT (\Mm_\text{s-model})^{-1}\mathbf{q}_m}\Big),\\
		p_\text{t-model}(\mathbf{Q}) = &\> \prod_m 
	\frac{1}{ (2\pi)^{N_m} \, \sqrt{ \det \Mm_\text{t-model} }}
	\exp\Big({-\frac{1}{2}\mathbf{q}_m^\TT (\Mm_\text{t-model})^{-1}\mathbf{q}_m}\Big),
\end{aligned}
\end{equation}
as functions of the unknown scatterer intensities~$\{\sigmasqalpha\}$ that enter~$\Mm$ via~\eqref{eq:Mjjp} for each of the models in~\eqref{eq:6nu}. 
The functions~$p_\text{s-model}(\mathbf{Q})$ and $p_\text{t-model}(\mathbf{Q})$ are called the likelihood functions~\cite{Mendenhall-73}. 
The discrimination procedure solves two optimization problems formulated as follows: 
\begin{equation}
\label{eq:ssttOptimization}
	\breve p_s = \max_{\sigmasqz,\sigmasqn,\sigmasqs} p_\text{s-model}(\mathbf{Q})
,\quad
	\breve p_t = \max_{\sigmasqz,\sigmasqn,\sigmasqt} p_\text{t-model}(\mathbf{Q})
	,
\end{equation}
subject to $\sigmasqz,\sigmasqn,\sigmasqs,\sigmasqt \ge 0$. The resulting $\breve p_s$ and $\breve p_t$ yield the maximum likelihood (ML) values for the corresponding scatterer models. It is common to consider the logarithm of the likelihood rather than the likelihood itself. Accordingly, we introduce 
\begin{equation}
\label{eq:ldef}
	l \bydef \log \breve p_t - \log \breve p_s
	, 
\end{equation}
and the classification based on the comparison of the two maxima \cite{a85e} is performed as follows: 
\begin{equation}
\label{eq:stAlgorithm}
\begin{array}{l}
	\text{\tt if $l > 0$}
\\
	\text{\tt then}
\\
	\quad \text{the target  is classified as a delayed scatterer (\ref{eq:6nut}), (\ref{eq:16Ityt})} 
\\
	\text{\tt else}
\\
	\quad \text{the target  is classified as an instantaneous scatterer (\ref{eq:6nus}), (\ref{eq:16Itys}).} 
\end{array}
\end{equation}

\section{Statistical characterization of observations}
\label{sec:allstats}

\subsection{Classification outcomes and confusion matrices}
\label{sec:twomatrices}

The results of classification by means of algorithm~\eqref{eq:stAlgorithm} may turn out incorrect for two different reasons. First, the outcome of algorithm~\eqref{eq:stAlgorithm} depends on the difference between the values of $\breve p_s$ and~$\breve p_t$ that are subject to computational errors and noise. For example, the classification decision based on a small value of~$|l|$, see~\eqref{eq:ldef}, should be considered unreliable.  
At the same time, a large positive value of~$l$ obtained from an individual image may give a strong indication that the underlying target is described by a t-model, i.e., has a delayed component. 

An extension of algorithm~\eqref{eq:stAlgorithm} that recognizes the issue of small values of~$|l|$ may look as follows: 
\begin{equation}
\label{eq:stAlgorithm_conf}
\begin{array}{l}
	\text{\tt if $ l > l^+$}
\\
	\text{\tt then}
\\
	\quad \text{the target  is classified as a delayed scatterer (\ref{eq:6nut}), (\ref{eq:16Ityt})} 
\\
	\text{\tt elseif $ l < l^-$}
\\
	\text{\tt then}
\\
	\quad \text{the target  is classified as an instantaneous scatterer (\ref{eq:6nus}), (\ref{eq:16Itys})} 
\\
	\text{\tt else}
\\
	\quad \text{the classification result is uncertain.} 
\end{array}
\end{equation} 
As compared to algorithm~\eqref{eq:stAlgorithm}, we have introduced two classification thresholds, $l^-$ and~$l^+$, to be defined in Section~\ref{sec:cdfs}, instead of a single threshold $l=0$. Accordingly, we have three classification outcomes: s-model, t-model, and uncertain, instead of the two outcomes in algorithm~\eqref{eq:stAlgorithm}. 

The second fundamental reason for possible misclassification is that 
formulae~\eqref{eq:pQ} yield a nonzero probability density for any model and any data~$\mathbf{Q}$, so a certain fraction of errors is inevitable regardless of the algorithm. The quality of the classification is characterized by the frequency of errors. 
Suppose that we have obtained a representative ensemble of sampled images of a target described by the s-model and another such ensemble for the t-model. Executing either of the algorithms \eqref{eq:stAlgorithm} or \eqref{eq:stAlgorithm_conf} on each image in these ensembles, we can evaluate the performance of the classification by means of the confusion matrices as in Table~\ref{table:confusionmatrices}. The rows named \mbox{``input: s''} and~\mbox{``input: t''} denote the models \eqref{eq:16Itys} and~\eqref{eq:16Ityt}, respectively, whereas the columns correspond to the outcomes of the particular classification algorithm. The ideal confusion matrix in Table~\ref{table:confusionmatrices}(a) will have $r_s=r_t=0$, whereas for Table~\ref{table:confusionmatrices}(b) this will be $r'_s=r'_t = r''_s=r''_t=0$. 

The frequency of classification errors depends on several factors. System parameters, e.g., bandwidth, aperture width, etc., form one group. Another group contains parameters of the target, such as its contrast. The roles of these groups of parameters have been investigated in~\cite{a85e}. Ultimately, the classification quality depends on the discrimination algorithm. The choice of the algorithm and its settings may depend on the specific application. For example, a wide gap between $l^-$ and~$l^+$ in~\eqref{eq:stAlgorithm_conf} should decrease $r'_s$ and $r'_t$ in Table~\ref{table:confusionmatrices}(b) at the cost of a large fraction of uncertain outcomes, i.e., large values of $r''_s$ and~$r''_t$. In Section~\ref{sec:allcontrasts}, we discuss a procedure whereby the classification errors can be kept below a specified level.

\begin{table}[ht!]
\caption{Confusion matrices: (a) for classification using algorithm~\eqref{eq:stAlgorithm}; (b) for classification using algorithm~\eqref{eq:stAlgorithm_conf}. The rows correspond to the models in~\eqref{eq:16Ity}, whereas the columns indicate the outcomes of a particular classification algorithm. The entries are relative frequencies of the events calculated for two ensembles with the same contrast, i.e., $q_s = q_t = q$. }
\label{table:confusionmatrices}
\begin{center}
\begin{tabular}{c l}
{\Large (a)} \hspace{0.5cm} & 
\begin{tabular} { | c | c | c |  }
\hline
	& {\bf output: s} & {\bf output: t} 
\\ \hline
	{\bf input: s} & $1 - r_s  $ & $r_s$ 	
\\ \hline
	{\bf input: t} & $r_t$ & $1 - r_t  $ 
\\ \hline 
\end{tabular} 
\\ \ \\[4mm]
{\Large (b)} \hspace{0.5cm} & 
\begin{tabular} { | c | c | c | c | } 
\hline
	& {\bf output: s} & {\bf output: t}  & {\bf uncertain} 
\\ \hline
	{\bf input: s} & $1 - r'_s - r''_s $ & $r'_s$ & $r''_s$  	
\\ \hline
	{\bf input: t} & $r'_t$ & $1 - r'_t - r''_t  $ & $r''_t$  
\\ \hline 
\end{tabular} 
\end{tabular}
\end{center}

\end{table}

\subsection{Confidence levels for classification with a given target contrast}
\label{sec:cdfs}

A standard approach to controlling the estimation errors for noisy measurements includes confidence intervals or levels~\cite{Mendenhall-73}. 
In parameter estimation problems, a confidence interval is built around the measured value of a certain parameter to indicate a possible range for the true value of this parameter. The boundaries of such an interval are determined from an ensemble of measurements of the parameter of interest or a probability distribution function representing it, such that only a small percentage of outliers, say 5\%, falls beyond this interval. Similarly, for a classification problem, an individual measurement can be assigned a numerical characteristic that will express the certainty that this observation falls into (or beyond) a specific category \cite{Mendenhall-73}. 
For the procedure described in Section~\ref{sec:stRadio_detection}, 
the value of~$l$ defined by~\eqref{eq:ldef} can play the role of such parameter. 

Yet in the case of SAR imaging, building an ensemble of observations to study the statistical properties of the discrimination procedure is not realistic, as explained in Section~\ref{sec:stRadio_Introduction}. In \cite{a85e}, we introduced a Monte-Carlo procedure that simulates ensembles of sampled coordinate-delay SAR images of instantaneous and delayed  targets, see (\ref{eq:16Ity}).%
\footnote{
To minimize the computational cost, we always choose $N_m=2$, $\psi_{m1} = \zetad$, $\psi_{m2} = -\zetad$, see~\eqref{eq:sumcalA}. Referring to Fig.~\ref{fig:noisyIsIt}, it means that for each~$\xid$, we sample a pair of coordinate-delay ``points''  
$\big(t_y=0,\; y_2=\zdT + \xid \cdot c/(B\sin\theta)\big)$ and 
$\big(t_y = \xid \cdot 2/B, \; y_2=\zdT\big)$. 
Choosing these two locations on a given ambiguity line has the advantage of maximizing the expectation of the intensity of at least one of the two possible inhomogeneous images, $|I_s|^2$ or $|I_t|^2$, see Fig.~\ref{fig:IsIt}. This is beneficial in the presence of  fluctuations due to the background and noise. 
}
We used those ensembles to evaluate the efficiency of algorithm~\eqref{eq:stAlgorithm} for different target contrasts. In the current work, we extend the approach of \cite{a85e} to define the confidence levels for target classification. 

%
\begin{figure}[ht!]
\hspace{-0.7in}
	\includegraphics[width=6.5in]{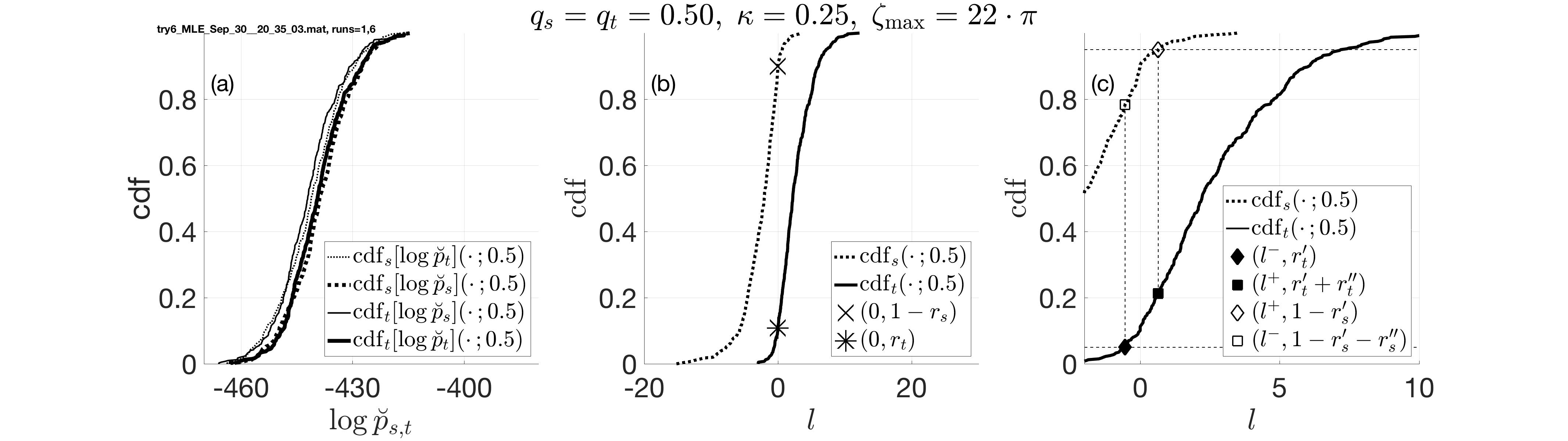}
\caption{
Cumulative distribution functions (cdf) for ensembles generated from the s-model and t-model.  
(a) Examples of cdfs for $\log \breve p_s$ and $\log \breve p_t$, see~\eqref{eq:ssttOptimization}.
The notations $\cdf_s[\ldots]$ and~$\cdf_t[\ldots]$ are similar to those in \eqref{eq:cdf2params}. 
(b) Examples of cdfs for  $l = \log \breve p_t - \log \breve p_s$, see \eqref{eq:ldef} and~\eqref{eq:cdf2params}. According to~\eqref{eq:alphabetacdf} (see also Table~\ref{table:confusionmatrices}(a)), we have 
$r_t = \cdf_t(0; q)$ and $r_s=1-	\cdf_s(0; q)$.
(c) A zoom-in to the central part of panel~(b). The entries $r'_s$, $r''_s$, $r'_t$, and~$r''_t$ from  Table~\ref{table:confusionmatrices}(b) are determined according to \eqref{eq:lplusminusdef} and~\eqref{eq:rppsrppt} with $p=0.05$. 
}
\label{fig:threecdfpanels}
\end{figure}

In the simplest setting, the simulated ensembles of sampled SAR images represent two scenarios in~\eqref{eq:16Ity} with equal target contrasts (\ref{eq:contrastdef}):
\begin{equation}
\label{eq:qsqtq}
	q_s=q_t = q
	,
\end{equation}
In addition to  contrast, each scenario has a set of associated parameters, such as $\kappa$, the values of $\{\zeta_m\}$ used for sampling, etc. The output of simulation is an ensemble of datasets~$\mathbf{Q}$ of type \eqref{eq:datasetQ} that we use in lieu of the actual measurements. 
While the discrimination procedure does not ``know'' which of the two target models in~\eqref{eq:16Ity} and what contrast were used to generate a given dataset~$\mathbf{Q}$, we can associate the outcomes of the procedure, and in particular, the set of values of $\breve p_s$ and~$\breve p_t$, with the type of underlying model and the values of its parameters.  
We will describe the statistics of these outcomes with the help of a cumulative distribution function~(cdf),%
\footnote{A more commonly used probability density function (pdf) is the first derivative of~cdf. 
}
which for a real-valued random variable~$\xi$ and a given argument $x$ yields the probability that $\xi<x$:
\begin{equation}
\label{eq:cdfdef}
	\cdf[\xi](x) \bydef P(\xi<x).
\end{equation}
In the context of discrimination between the two types of scatterers, the random variable will be~$l$ defined in~\eqref{eq:ldef}, and we will use the following notations:
\begin{subequations}
\label{eq:cdf2params}
\begin{equation}
\label{eq:cdf2params_s}
	\cdf_s(x;q) \bydef \cdf[l](x) 
\end{equation}
for the ensemble generated from the s-model with $q_s=q$, and
\begin{equation}
\label{eq:cdf2params_t}
	\cdf_t(x;q) \bydef \cdf[l](x) 
\end{equation} 
\end{subequations}
for the ensemble generated from the t-model with $q_t=q$. 
While the target contrast is explicitly specified as the second argument of $\cdf$ in~\eqref{eq:cdf2params}, other system and target parameters affecting the probability in~\eqref{eq:cdfdef} will be considered fixed until Section~\ref{sec:stRadio_results}. 
Note that the subscript at a $\cdf$ in (\ref{eq:cdf2params}) corresponds to the {\em rows} in the confusion matrices in Table~\ref{table:confusionmatrices}, whereas the choice of the model in the optimization problem is denoted by the lower index in~$\breve p_s$ and $\breve p_t$, see \eqref{eq:pQ} and~\eqref{eq:ssttOptimization}. 

Figure~\ref{fig:threecdfpanels}(a) plots cdfs of $\log \breve p_s$ and $\log \breve p_t$ for a pair of ensembles of sampled images that differ only by the type of the actual inhomogeneous scatterer. For the same data, 
Figure~\ref{fig:threecdfpanels}(b) plots $\cdf_s(\cdot,q)$ and $\cdf_t(\cdot,q)$, see~\eqref{eq:cdf2params}, for $q_s=q_t=q=0.5$. 
This plot clearly shows the separation between these two ensembles, such that most of the values of~$l$ are negative for the ensemble generated from the s-model and positive for the ensemble generated from the t-model. This implies that the discrimination results by algorithm~\eqref{eq:stAlgorithm} are correct in most cases (remember that $l$~is calculated from the observations by a procedure that has no access to the underlying value of contrast or model type). 
We can establish the following relation between the curves  in Fig.~\ref{fig:threecdfpanels}(b) and the values in Table~\ref{table:confusionmatrices}: 
\begin{subequations}
\label{eq:alphabetacdf}
\begin{align}
\label{eq:alphabetacdf_t}
	\cdf_t(0; q) &\> = r_t\\
\intertext{and, similarly,}  
\label{eq:alphabetacdf_s}
	\cdf_s(0; q) &\> = 1-r_s.
\end{align}
\end{subequations}
For example, the value of~$\cdf_t(0;q)$ yields the fraction of targets in the ensemble built from the t-model that algorithm~\eqref{eq:stAlgorithm} incorrectly classifies as s-targets.


To introduce confidence levels, we choose a small value, say $p=0.05$, i.e., 5\%,
as a threshold for admissible classification errors. In other words, our goal is to make sure that 
\begin{equation}
\label{eq:rpsrpt_goal}
	r'_s \le p \quad \text{and} \quad r'_t \le p, 
\end{equation}
see Table~\ref{table:confusionmatrices}(b). 
Then, we define two values, $l^-$ and~$l^+$, implicitly as solutions to the following equations:
\begin{equation}
\label{eq:lplusminusdef}
	\cdf_t(l^-;  q) = p
\quad \text{and}\quad 
	\cdf_s(l^+; q) = 1-p
	. 
\end{equation}
The cdfs in~\eqref{eq:lplusminusdef} are nondecreasing in their first argument, but may be discontinuous. Although this does not present a major obstacle to  subsequent considerations, we will assume for simplicity that all cdfs are continuous and monotonic; in this case, solutions $l^-$ and~$l^+$ always exist and unique for $p<1$. 

We will consider first the case where $l^- < l^+$, which is equivalent to 
\begin{equation*}
	\cdf_s(l^-; q) - \cdf_t(l^-; q) < 1-2p
,\quad 
	\cdf_s(l^+; q) - \cdf_t(l^+; q) < 1-2p
	,
\end{equation*} 
as shown in Fig.~\ref{fig:threecdfpanels}(c) for $p=0.05$. 
For an ensemble of datasets~$\mathbf{Q}$ generated from the t-model, the frequency of the cases $l(\mathbf{Q}) < l^-$ will be equal to~$p$. If this value of~$l^-$ is used as the lower threshold in algorithm~\eqref{eq:stAlgorithm_conf}, with the above dataset as the input, we will also have 
\begin{subequations}
\label{eq:rpsrpt}
\begin{equation}
	r'_t = 
	P\big(l(\mathbf{Q}) < l^- \; | \; \text{t-model} \big) = p,  
\end{equation}	
Considering an ensemble generated from the s-model, we obtain, in a similar way, the following: 
\begin{equation}
	r'_s = 
	P\big(l(\mathbf{Q}) \ge l^+ \; | \; \text{s-model} \big) = p. 
\end{equation}
\end{subequations}
From relations~\eqref{eq:rpsrpt}, we see that using the interval $(l^-,l^+)$ defined by~\eqref{eq:lplusminusdef} in algorithm~\eqref{eq:stAlgorithm_conf}, we can keep the rate of classification errors, in particular, $r'_s$ and~$r'_t$ in Table~\ref{table:confusionmatrices}(b), at the predefined level as stated in~\eqref{eq:rpsrpt_goal}. 

\begin{figure}[ht!]
\hspace{-0.5in}
	\includegraphics[width=6in]{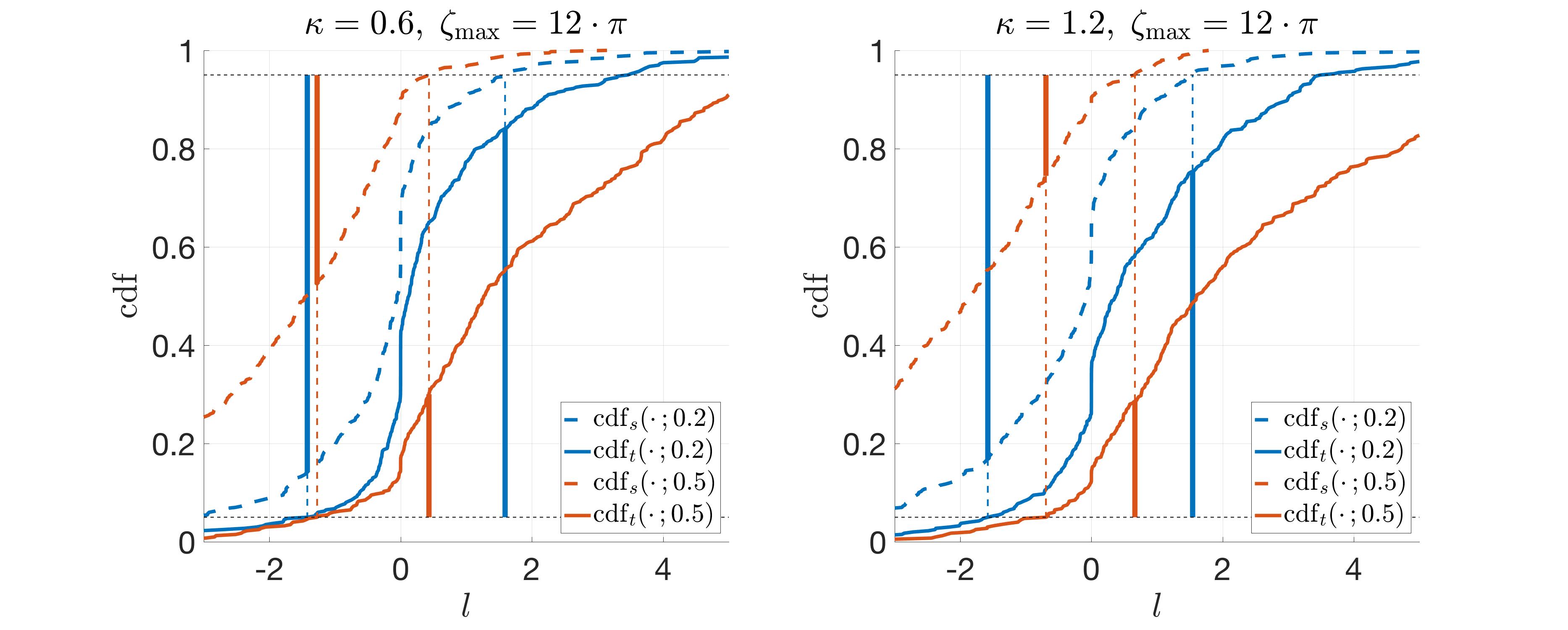}
\vspace{-0.2in}
\caption{
Separation between the graphs of $\cdf_s(\cdot\,;q)$ and~$\cdf_t(\cdot\,;q)$ for different values of $\kappa$ and~$q$. The thick colored vertical bars indicate the percentage of uncertain classifications for the ensembles generated from the s-model (the left set of bars in each plot) and t-model (the right set of bars), see also \eqref{eq:rppsrppt} and Fig.~\ref{fig:threecdfpanels}(c). 
}

\label{fig:undecided_bars}
\end{figure}

The rate of uncertain outcomes from algorithm~\eqref{eq:stAlgorithm_conf} can be expressed as follows: 
\begin{equation}
\label{eq:rppsrppt} 
\begin{aligned}
	r''_t 
&\bydef\> 
	\cdf_t(l^+;q) - \cdf_t(l^-;q)
         = \cdf_t(l^+;q) - p, 
\\
	r''_s 
&\bydef\> 
	\cdf_s(l^+;q) - \cdf_s(l^-;q)
	= 1 - p - \cdf_s(l^-;q).                
\end{aligned}
\end{equation}
Relations \eqref{eq:lplusminusdef} and~\eqref{eq:rppsrppt} are illustrated in Figs. \ref{fig:threecdfpanels}(c) and~\ref{fig:undecided_bars}. 

The case where from~\eqref{eq:lplusminusdef} we obtain $l^- \ge l^+$ (e.g., for the ensembles in Fig.~\ref{fig:threecdfpanels}(c), this will happen for~$p=0.2$) can be interpreted as follows: the separation  between the two ensembles of values of~$l$ is so good that the error level of~$p$ can be guaranteed with no need for a confidence interval. In this case, we can use algorithm~\eqref{eq:stAlgorithm} with any $l^* \in [l^+,l^-]$ as a single threshold. Alternatively, we can find~$l^*$ as a solution to 
$$
	\cdf_s(l^*,q) + \cdf_t(l^*,q) = 1.  
$$
As the cdfs are monotonic, we will have $l^* \in [l^+,l^-]$ yielding the error rates of algorithm~\eqref{eq:stAlgorithm} at
$$
	r_t  = \cdf_t(l^*,q) = 1-\cdf_s(l^*,q)  = r_s \le p, 
$$
which also satisfies~\eqref{eq:rpsrpt_goal}. 

%

Finally, we should note that taken alone, the definitions of thresholds in~\eqref{eq:lplusminusdef} can be seen as a way of excluding either extremely large positive or extremely large negative values of~$l$. However, when the thresholds defined in~\eqref{eq:lplusminusdef} are used in algorithm~\eqref{eq:stAlgorithm_conf}, it is a neighborhood of $l=0$ that gets thrown away. This highlights the difference between the problems of parameter evaluation and classification. For the latter, once the rate of classification errors has been fixed at~$p$, see~\eqref{eq:rpsrpt}, the quality of classification is determined by the percentage of uncertain outcomes, i.e., the values of $r''_s$ and~$r''_t$ in Table~\ref{table:confusionmatrices}(b). From Fig.~\ref{fig:undecided_bars}, we can see that as either~$\kappa$ or~$q$ increase, the curves of $\cdf_s(\cdot\,;q)$ and~$\cdf_t(\cdot\,;q)$ become better separated, the intervals between $l^-$ and~$l^+$ shrink, and the above percentages decrease, as expected. 

\subsection{Generalization to all target contrasts} 
\label{sec:allcontrasts}

The confidence intervals introduced in Section~\ref{sec:cdfs} depend on the target contrasts $q_s$ and~$q_t$ defined by~\eqref{eq:contrastdef}. The latter values should be considered unavailable to the image processing algorithm. Hence, the definitions of $l^-$ and~$l^+$ in~\eqref{eq:lplusminusdef} should be modified in order to make them independent of target contrasts. 

One way of achieving this goal is to use prior information about the target contrasts. For example, we can assume that the probability distribution of the target contrast is known. This means that we can consider~$q$ to be a random variable with known probability, and  instead of ensembles with a given value of~$q$ used in Section~\ref{sec:cdfs} generate a pair of ensembles, one for the s-model and one for the t-model, with the given statistics of target contrasts (in general, this statistics can be different for s-target and t-target models). Then, $\cdf_s$ and~$\cdf_t$ built from these ensembles should replace the cdfs of~\eqref{eq:cdf2params} in definitions~\eqref{eq:lplusminusdef}. 
 
An alternative approach that uses no prior information about the contrast is to take the minimal~$l^-$ and, correspondingly, maximal~$l^+$, over the entire range of target contrasts: 
\begin{equation}
\label{eq:gen_lplusminus}
	l^-=\min_q \big(x \; | \cdf_t(x; q) = p \big) 
,\quad
	l^+=\max_q\big(x \; | \cdf_s(x; q) = 1-p \big)
	.
\end{equation}
With $l^-$ and~$l^+$ redefined as in~\eqref{eq:gen_lplusminus}, 
the procedure~\eqref{eq:stAlgorithm_conf} should perform with the classification error rates $r'_s$ and~$r'_t$ not exceeding~$p$ for ensembles generated from any probability distributions of contrasts~$q$. 

Definitions~\eqref{eq:gen_lplusminus} are used in the numerical experiments described in Section~\ref{sec:stRadio_results}. In order to actually compute the maximum and minimum in~\eqref{eq:gen_lplusminus}, we have taken the values of~$q$ in the range from~$0.0$ to~$0.9$ with the step of~$0.1$.

\section{Simulation results}
\label{sec:stRadio_results}


Discrimination between the instantaneous and delayed targets hinges upon our ability to resolve the range-delay ambiguity (see equations \eqref{eq:ambig_instant} and~\eqref{eq:ratioOfIs}) in the presence of clutter and noise. 
The quality of discrimination depends on the system and target parameters. In Section~\ref{sec:allstats}, the only variable parameter of the model was the contrast~$q$ of~\eqref{eq:qsqtq}. In this section, we  explore the dependence of the discrimination quality on the parameters~$\kappa$ and~$\zetamax$ that determine the threshold for having the range-delay ambiguity resolved, see \eqref{eq:kz20}.

\begin{figure}[ht!]
\begin{center}
	\includegraphics[width=5.5in]{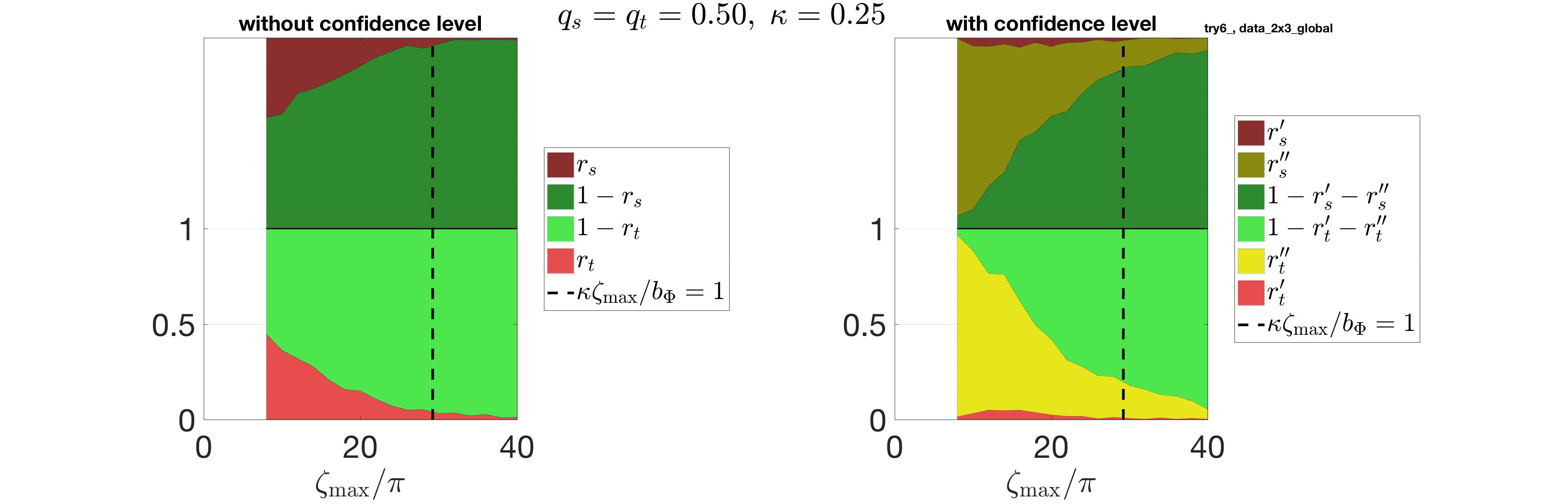}
	\includegraphics[width=5.5in]{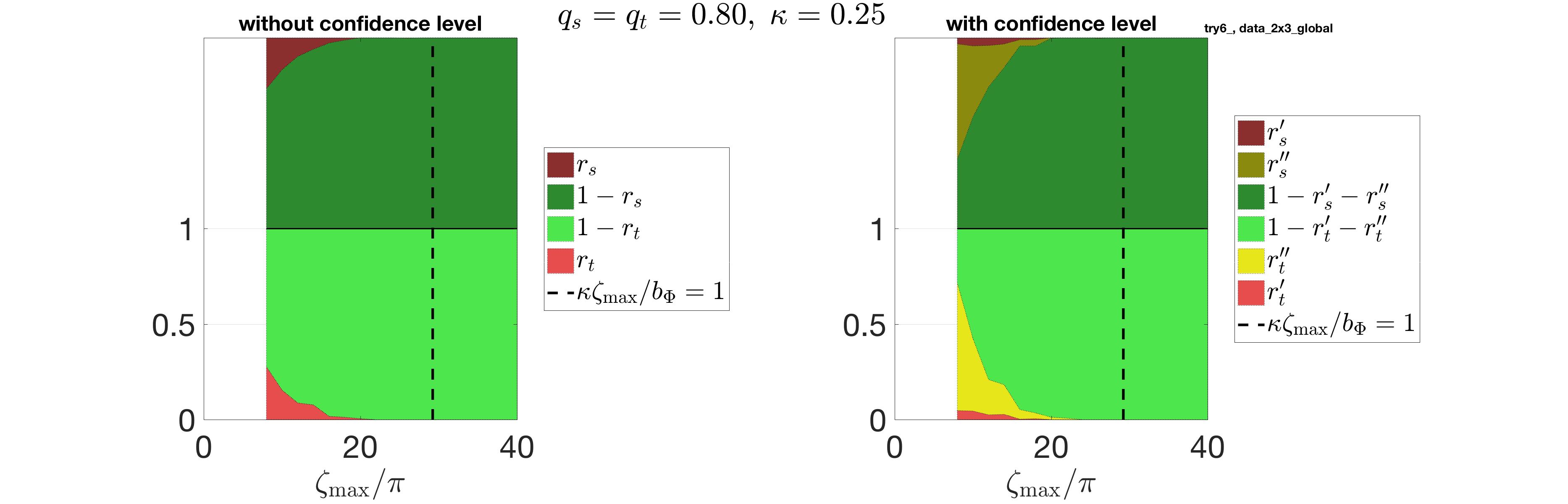}
\vspace{-0.2cm}
\end{center}
\caption{Dependence of the discrimination quality on~$\zetamax$, see~\eqref{eq:finitesupportF}, for two different target contrasts. The left column of plots corresponds to algorithm~\eqref{eq:stAlgorithm} and Table~\ref{table:confusionmatrices}(a), and the right column corresponds to algorithm~\eqref{eq:stAlgorithm_conf} and Table~\ref{table:confusionmatrices}(b). 
The dashed vertical lines are drawn at $\kappa\zetamax = \bPhi \approx 23$, cf.~\eqref{eq:kz20}. 
The lower and upper parts of the colored panels represent ensembles generated from the s-model and t-model, respectively. The percentage of correct classifications is shown in green (two different shades are used to distinguish between the ensemble types), incorrect in red, and uncertain in yellow (only the right column of plots).
}
\label{fig:zetamax_3colors}
\end{figure}

\begin{figure}[ht!]
\begin{center}
	\includegraphics[width=5.5in]{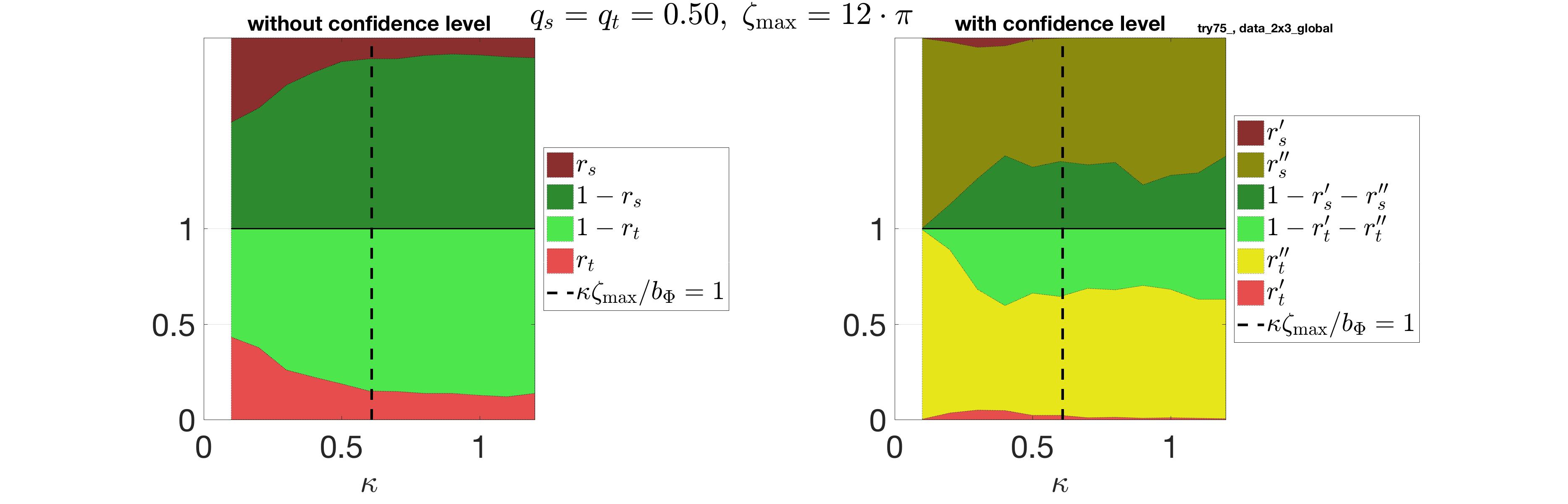}
	\includegraphics[width=5.5in]{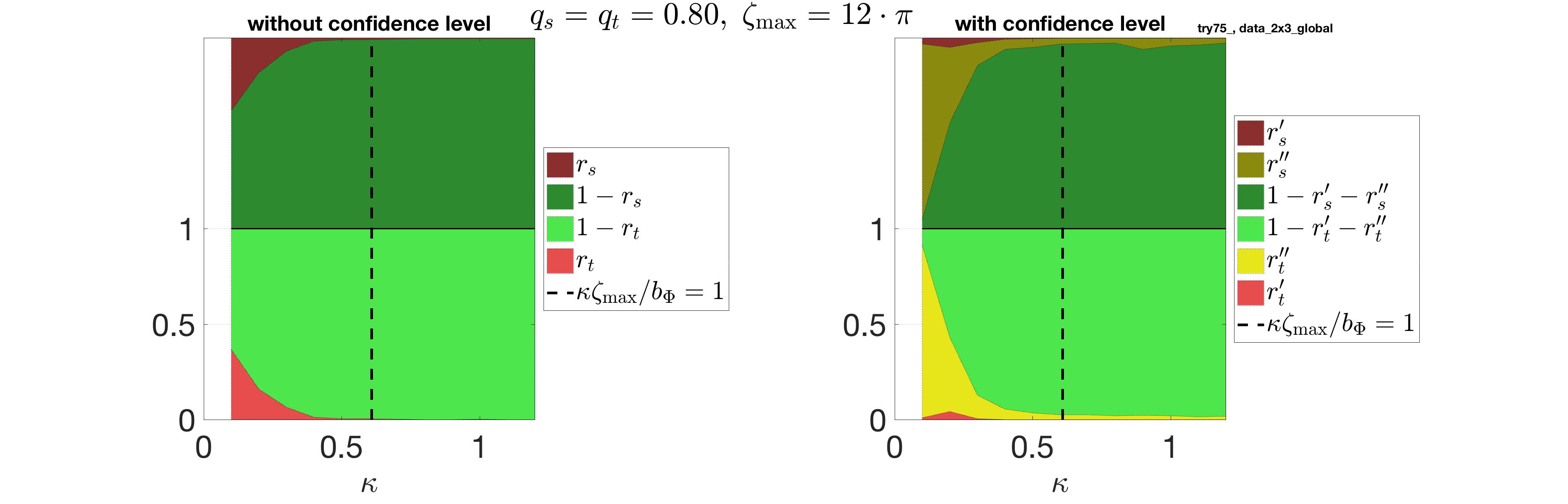}
\vspace{-0.2cm}
\end{center}
\caption{Dependence of discrimination quality on~$\kappa$, see~\eqref{eq:kappadef}. The notations are the same as in Fig.~\ref{fig:zetamax_3colors}. 
}
\label{fig:kappa_3colors}
\end{figure}

Figs.\ \ref{fig:zetamax_3colors} and~\ref{fig:kappa_3colors} show the dependence of the off-diagonal entries of the confusion matrices in Table~\ref{table:confusionmatrices}    
on $\zetamax$ and~$\kappa$, respectively, for two different values of the target contrast $q$. 
The lower half in each  color panel represents the second row in Table~\ref{table:confusionmatrices}(a) or \ref{table:confusionmatrices}(b), with the colors denoting the individual entries. The upper half corresponds to the first rows in Table~\ref{table:confusionmatrices}(a) or \ref{table:confusionmatrices}(b); for clarity of presentation, this part is flipped vertically with respect to the bottom half. 
The dashed vertical lines are drawn at $\kappa\zetamax = \bPhi$, where $\bPhi \approx 23$ \cite{a85e} is the first local minimum of~$|\Phi(0,\cdot)|$, see \eqref{eq:Phiv2} and \eqref{eq:kz20}. 

As expected, the discrimination quality improves with the increase of~$\zetamax$, see Fig.~\ref{fig:zetamax_3colors}. A less expected effect that can be observed in Fig.~\ref{fig:kappa_3colors} is the saturation of the fraction of reliable classifications for $\kappa \gtrsim 0.4$; this may require further attention.  
Introduction of the confidence level successfully keeps the number of classification errors below~$p$. This, on the other hand, makes a  number of correct classifications deemed uncertain. 

\section{Discussion}
\label{sec:stRadio_discussion}

The goal of analyzing the scattering delay is to enhance the amount of information supplied by a radar imaging system as compared to standard SAR. At the same time, the proposed methodology uses  tools from  image classification and pattern recognition. Hence, future developments of this work may come from solving two completely different classes of problems. 

In the field of radar imaging, one possibility for the next step is to consider a wider class of functions~$F_s$ and~$F_t$ as compared to the characteristic  functions~\eqref{eq:finitesupportF} used in this work. Another option is to explore the stability of the discrimination method to the incorporation of highly coherent components in the received signal: this problem was addressed by means of time-frequency analysis in~\cite{borden-98,chen-2002b}. Additional steps that can improve the applicability and performance of the discrimination procedure are suggested in~\cite{a85e}. Further, the choice of the contrast parameter~$q$ for setting the confidence levels, see~\eqref{eq:qsqtq} and~\eqref{eq:gen_lplusminus}, may not always be optimal from the standpoint of applications. For example, in a different setting we may be interested in detecting the cases where the dimensionless delay $\zetamax$ of~\eqref{eq:finitesupportF} exceeds a certain threshold value. Such a problem will require significant modification to the classification algorithm~\eqref{eq:stAlgorithm_conf} and definitions of the confidence levels~\eqref{eq:gen_lplusminus}. 

The most noticeable developments in the area of image classification and pattern recognition are currently related to the advances in the artificial intelligence (AI) \cite{bishop-2006}. The concepts of deep learning and multi-layer convolution neural network (CNN) have received a wide recognition because of their demonstrated efficiency in image classification tasks \cite{krizhevsky-2012,mishkin-2017}. Yet introducing elements of AI into the analysis of coordinate-delay SAR images may be complicated for several reasons. First, these  images are expensive to build, and we cannot expect to be able to obtain the training sets as massive as those with  optical images. Second, the appearance and properties of ``signal'' and ``noise'' in coordinate-delay SAR images, see Fig.~\ref{fig:noisyIsIt}, are very different from those in photography. Hence, besides the convolution and activation  operations (i.e., nonlinearity) that are the building blocks of an image classification CNN, we may want to use the transformations that take into account the correlation properties of the images given by~\eqref{eq:psidpsidp}--\eqref{eq:HbHtHs}. 
Reports about successful application of deep learning to the problems of automated target recognition in standard SAR images are encouraging  \cite{zhu-2017,chen2014sar}, but at the same time the scarcity of the real data and the difficulties in augmenting it with modelled data are recognized as a major problem \cite{chen2014sar,scarnati-2019-deep}. 

As a combination of these two directions, we can apply the modern classification techniques to the entire output of the optimization problems~\eqref{eq:ssttOptimization}. This means that in addition to the minimum values used in the classifier~\eqref{eq:stAlgorithm}, we will take into account the arguments of the minima, i.e., the minimizing scatterer intensities: 
$
	(\breve p; \breve{\sigma}_\alpha^2, \alpha \in {\cal S})_\text{s-model} \times 
	(\breve p; \breve{\sigma}_\alpha^2, \alpha \in {\cal T})_\text{t-model}
	, 
$
see~\eqref{eq:calA} and~\eqref{eq:ssttOptimization}.	
The resulting parameter space is 8-dimensional, which is hard to process without assistance from some classification algorithm. In our initial trials involving a linear classifier (see, e.g., \cite[Chapter~4]{hastie-2009}), we did not observe any significant improvements as compared to the method~\eqref{eq:stAlgorithm} that uses only two out of the eight parameters. This topic may require more attention in the future. 

\section*{Acknowledgements}

We are grateful to Profs.\ Alen Alexanderian and Ralph Smith (NCSU) for fruitful discussions.  
This material is based upon work supported by the US Air Force Office of Scientific Research (AFOSR) under award number FA9550-17-1-0230.

\bibliography{../references}
\bibliographystyle{unsrt}

\end{document}